\documentclass[11pt,a4paper]{article}

\usepackage[utf8]{inputenc}
\usepackage{amsfonts}
\usepackage{a4wide}
\usepackage{caption}
\usepackage{subcaption}
\usepackage{siunitx}
\usepackage{booktabs}
\usepackage{microtype}
\usepackage{lmodern}
\usepackage{algorithm}
\usepackage{algorithmicx}
\usepackage{algpseudocode}
\usepackage{authblk}

\usepackage{float}
\usepackage{colortbl}
\usepackage{amsmath,amssymb}
\usepackage{dsfont}
\usepackage{hyperref}
\usepackage{graphicx}
\usepackage{enumitem}
\usepackage{arydshln}
\usepackage{mathtools}
\usepackage{bbold}
\usepackage{soul}
\usepackage{indentfirst}
\usepackage[dvipsnames]{xcolor}

\newcommand\mumu{\ensuremath{{\mu^+\mu^-}}}

\newcommand\Bz{\ensuremath{{B^0}}}
\newcommand\Kstarz{\ensuremath{{K^{*0}}}}
\newcommand\jpsi{\ensuremath{{J\!/\!\psi}}}
\newcommand\psitwos{\ensuremath{{\psi(2S)}}}

\begin{document}

\title{\textbf{The DL Advocate: Playing the devil's advocate with hidden systematic uncertainties}\vspace{5mm}}

\author[1]{Andrei Golutvin}
\affil[1]{Imperial College of London, London, United Kingdom}

\author[2]{Aleksandr Iniukhin}
\affil[2]{Yandex School of Data Analysis}

\author[1]{Andrea Mauri}

\author[3]{Patrick Owen}

\author[3]{Nicola Serra}
\affil[3]{Physik-Institut, Universit\"at Z\"urich, CH-8057 Z\"urich, Switzerland}

\author[4,5]{Andrey Ustyuzhanin}
\affil[4]{Constructor University, Campus Ring 1, 28759 Bremen, Germany}
\affil[5]{Institute for Functional Intelligent Materials, National University of Singapore, Singapore 117544, Singapore}


\maketitle

\begin{abstract}
\noindent
We propose a new method based on machine learning to \emph{play the devil's advocate} and investigate the impact of unknown systematic effects in a quantitative way. This method proceeds by reversing the measurement process and using the physics results to interpret systematic effects under the Standard Model hypothesis.
We explore this idea with two alternative approaches: the first one relies on a combination of gradient descent and optimisation techniques, its application and potentiality is illustrated with an example that studies the branching fraction measurement of a heavy-flavour decay.
The second method employs reinforcement learning and it is applied to the determination of the $P_{5}^{'}$ angular observable in $B^0 \to K^{*0} \mumu$ decays.
We find that for the former, the size of a hypothetical hidden systematic uncertainty strongly depends on the kinematic overlap between the signal and normalisation channel, while the latter is very robust against possible mismodellings of the efficiency.
\end{abstract}

\clearpage

\section{Introduction}

The potential for hidden systematic effects, so-called "unknown-unknowns", in a physics measurement is difficult to address. This problem can be alleviated by an independent confirmation in a different experiment. A good example, in particle physics, is the discovery of the Higgs boson, whereby the simultaneous announcement from both the ATLAS and CMS experiments of an excess led to high confidence in the discovery \cite{Aad_2012,Chatrchyan_2012}. However, the size of the collaborations and the complexity of the experiments involved can make such independent confirmations prohibitively expensive for future particle physics experiments. The confidence in the Higgs discovery was also aided by the fact that it was not completely unexpected: it was predicted by the Standard Model (SM) of particle physics to emerge with a distinctive pattern of couplings to the known particles.  

The discovery of physics beyond the SM is expected to answer many of the open questions of the SM and it is therefore the main focus of experimental particle physics.  This discovery will require even more experimental evidence to be confirmed, particularly if it manifests itself in ways that are unexpected. The question in this case is therefore: under which conditions can one claim a physics discovery in an experiment which has unique physics sensitivity and therefore no direct competitors? The answer to this question is normally qualitative, such as seeing a particular control channel pass a set of compatibility tests or observe new physics appear with multiple and complementary experimental signatures. The goal of this paper is to introduce a method that provides quantitative answers to these questions.

The philosophy this paper follows is to apply deep learning techniques  to \emph{play the devil's advocate} (DL Advocate), with respect to deviations from the SM. This implies assuming the true value of the parameters of interest is indeed SM-like and determine what experimental effects could cause the observed set of measurements.
Concretely, one can consider the set of measurements of the experiment as a system of equations:

\begin{equation}
   {\cal F}(\tilde{\eta}, \Omega_i)=M_i\Longrightarrow 
   \begin{cases} 
   {\cal F}(\tilde{\eta}, \Omega_1)=M_1 \\
   {\cal F}(\tilde{\eta}, \Omega_2)=M_2 \\
   \cdots
   \label{eq:syst_eq_compact}
   \end{cases},
\end{equation}
where ${\cal F}$ represents the measurement process, $\tilde{\eta}$ are parameters in common to all measurements, such as the detector response, and $\Omega_i$ are parameters specific to a particular measurement, such as theoretical parameters. 
When a measurement deviates from its SM prediction, it is tempting to interpret the observed deviation as a sign of physics beyond the SM.
However, such an important claim must be supported by an equally strong confidence in the understanding of the experimental apparatus.
The idea developed in this paper reverses the classical reasoning and, instead of attributing the observed deviation to physics beyond the SM, it starts from the SM hypothesis by fixing the theory parameters to their predicted values,  uses the simulation to model ${\cal F}$ and uses a neural network to find possible values of $\tilde{\eta}$ that reproduce the observed measurements $M_i$.
In other words, it tries to find possible detector effects that can cause the observed deviation.
For the examples illustrated in this paper, the parameters $\tilde{\eta}$ represent a mismodelling of the detector efficiency, but can be extended to any assumption in the analysis. The system in Eq.~\ref{eq:syst_eq_compact} could in principle include all measurements by the experiment. However, in practice one can only consider measurements which are correlated either through theory or experiment.

The advantage of this approach is that the resulting values of $\tilde{\eta}$ can then be used to make quantitative predictions of mismodelling that can be falsified by additional crosschecks. There are two categories of information that $\tilde{\eta}$ can represent. One is \emph{high-level} information, such as kinematic information of particles. A mismodelling as a function of a particle momentum would fall into this class of information. Such information can be readily implemented with existing tools, but requires physics intuition and therefore has to be tuned to each specific case. The other category of information is \emph{low-level} quantities, such as the material budget and hit resolution. Exploring low-level information would require tuning simulation in real time and would be a challenge to implement, but would be fully general, applicable for any measurement that relies on simulation.

In recent years increasingly more interest has been devoted to apply machine learning techniques to systematic uncertainties. The main emphasis however has been on the optimisation of statistical inference in the presence of systematic uncertainties (see \emph{e.g.} ~\cite{Wunsch:2020iuh,DeCastro:2018psv,Ghosh:2021roe}) or incorporating known systematic effects in simulation (see \emph{e.g.}~\cite{Viren:2022qon,Englert:2018cfo,Louppe:2016ylz}). 
In this work we propose a method to bound the size of unknown or underestimated systematic effects.
A quantitative and incontrovertible demonstration of the control of systematic uncertainties is in fact essential for any scientific discovery, whose claim must be supported with a confidence higher than one-in-a-million chance not to be a fluctuation of statistical or systematic origin.

In order to demonstrate the potential of such an approach, we apply this method to a simple toy example of a branching fraction measurement of a particle decay and restrict our attention to a potential mis-modelling of the efficiency. Such measurements are often normalised to a decay mode with a known branching fraction and ideally the same final state as the signal, which cancels systematic uncertainties due to efficiency mismodelling to a high degree. It is therefore an ideal testing ground for our approach. We also consider a more concrete example which is the observable $P^\prime_{5}$, which arises in the angular distribution of $\Bz\to\Kstarz\mumu$ decays and is highly sensitive to physics beyond the SM~\cite{DescotesGenon:2012zf,Matias:2012xw}. In both cases we show in a simplified scenario the use of the DL Advocate technique to investigate the impact of possible efficiency mismodelling on the considered measurements.
Finally, another interesting system to test this methodology would be the W mass, recently been measured by the CDF collaboration~\cite{doi:10.1126/science.abk1781} to be significantly different from previous measurements~\cite{CDF:2013dpa,ATLAS:2017rzl,LHCb:2021bjt} and the Standard Model prediction~\cite{deBlas:2021wap}.

This paper is structured as follows:  The general idea and the algorithm implementation is described in Sect.~\ref{sec:method}.
The concrete example of the branching fraction measurement of a particle decay is briefly summarised in Sect.~\ref{sec:toy_example}. An implementation with Deep Reinforcement learning applied to the observable $P_{5}^{'}$ is discussed in Sect.~\ref{sec:RL}, which is followed by a summary section~\ref{sec:Discussion} and a conclusion.

\section{Methodology}
\label{sec:method}

Measurements in experimental particle physics typically involve the determination of a signal yield, either integrated over a particular decay channel or differentially in regions of phase-space.
Due to imperfections of the detection process, the recorded signal yield must be corrected for the finite detector efficiency in order to retrieve the  \emph{true} number of signal events originally produced.
This efficiency is typically estimated based on simulation or calibration samples.

Taking the example laid out in Eq.~\ref{eq:syst_eq_compact}, we consider a set of measurements $M_{i}$, where one is the measurement of interest and the others are control channels which are used to validate the analysis and to check for systematic uncertainties. For each of these measurements there are some observed candidates $N_i$ and an associated efficiency  $e_i$, so that $M_i=\frac{N_i}{e_i}$. 
The candidates of each channel are characterised by a set of variables (features) such as the kinematics of the produced particles. Differences in these distributions, together with a detection efficiency which can depend on the same kinematic variables, can result in different total efficiencies between the signal and control channels.

Broadly speaking, mismodelling of the efficiency and unaccounted or mismodelled backgrounds can create a bias in the measurements, which are accounted for by introducing ad-hoc systematic uncertainties.
As a consequence, partial or incorrect evaluation of these effects would result in underestimated systematic uncertainties. 
In this paper we will focus on the role of the efficiency.   

We describe possible mismodelling of the efficiency with a weighting function $w(x)$, which depends on the kinematic variables $x$ of the event. Values of $w(x)=1$ correspond to a perfect modelling of the efficiency, while values below/above unity correspond to efficiency under/over estimated. 
The key idea is that, while the detector response depends entirely on the kinematics of the single event, the total signal/control channel efficiency can suffer from different biases once integrated over the individual kinematic distribution of each decay channel.
We can then define the \emph{true} total efficiency for a given channel $i$ as 
\begin{equation}
e_i = \mathbb{E}_{x\sim p(x \vert i)}\left[w(x) \times \hat{\epsilon}(x)\right] \, ,
\end{equation}
where $\hat{\epsilon}(x)$ is the per-event \emph{estimated} efficiency in the experiment and the expectation value indicates the weighted average over the kinematic distribution of each decay channel $p(x \vert i)$.
Since our goal is to study the impact of possible mismodelling of the efficiency in a given set of measurements we can safely assume $\hat{\epsilon}(x) =1$ without loss of generality.
This simplifies the expression of the per-channel efficiency to
\begin{equation}\label{eq:eff_def}
e_i = \mathbb{E}_{x\sim p(x \vert i)}\left[w(x) \right] \equiv 
\frac{1}{n_i} \sum_{k=1}^{n_i} w(x_{k,i}) \, ,
\end{equation}
where we approximated the expectation value with a sum over a large number $n_i$ of simulated events $\{x_{k}\}_i$, where $i$ labels the different decay channels.

Control channels provide important constraints on how well the efficiency is estimated.
They are typically selected with topology and kinematics similar to the  signal decay mode in order to maximise the phase space overlap between channels.
The result of the measurements obtained on such control channels can then be compared to known reference values, \emph{e.g.} existing precise measurements from other experiments or clean SM predictions. 
If a good agreement is found, a certain level of confidence can be ascribed to the estimation of the efficiency, at least for what concerns the kinematic regions populated by the control channels.
The requirement that measurements performed on the control channels must be compatible with a certain reference can be formulated as 
\begin{equation}\label{eq:M_constr}
    M_{i}  \in [M_{i}^{low};M_{i}^{high}] 
\end{equation}
which reduces to
\begin{equation}\label{eq:eff_constr}
    e_{i}  \in [V_{i}^{low};V_{i}^{high}] \, ,
\end{equation}
once we restrict the attention to the sole role of the efficiency in the measurement.
Here $V_{i}^{low}$ and $V_{i}^{high}$ are the values that bound the efficiency for the control measurements to pass scrutiny.

The goal of this paper is to find regions of the kinematic space $\vec{x}$ where a mismodelling of the efficiency can have a significant  impact in the signal measurement while the effect on the control channels remains within the constraints of Eq.~\ref{eq:eff_constr}.
Given these constraints, the intuition is that this problem can be thought as a classification task between the signal and control channels using the kinematic variables provided in the space $\vec{x}$.

\subsection{DL Advocate algorithm}

\begin{figure}[t]
\centering
\includegraphics[width=0.65\columnwidth]{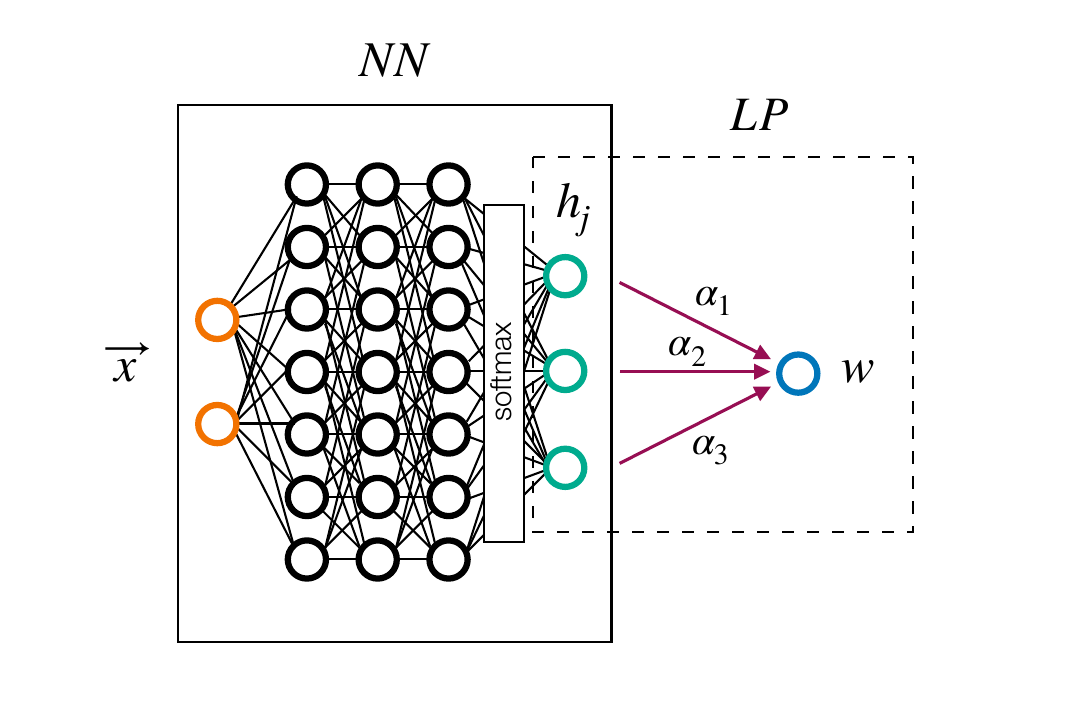}
    \caption{Schematic view of the DL Advocate algorithm. 
    The algorithm is formed by two parts: the output of a neural network (NN) is passed to a linear programming (LP) solver which returns the final weight for each event.}
    \label{fig:architecture}
\end{figure}

The algorithm is formed of two main parts, as shown schematically in Fig.~\ref{fig:architecture}. 
The first is a fully connected neural network which resembles a multi-classification algorithm. 
The inputs are a set of features $x$ whereas the output $h_j$ is a classification score for each decay hypothesis. 
The last layer of the NN is normalised with a softmax activation function which enforce $h_j(x) \geq 0$ and $\sum_j h_j(x) = 1$.
Details on the technical implementation of the neural network are given in App.~\ref{app:algorithm}.

The second part of the algorithm consists of a linear combination of the NN output, also referred to as linear programming (LP), which defines the final per-event weight
\begin{equation}\label{eq:w_def}
w(x) = \sum_j \alpha_j h_j(x)  \quad \mathrm{with} \quad \alpha_j \geq 0 \, .
\end{equation}
Combining this with Eq.~\ref{eq:eff_def} and moving to a vectorial notation we can express the total per-channel efficiency as 
\begin{equation}
    e_i =  \frac{1}{n_i} \sum_k  \vec{\alpha} \cdot \vec{h}(x_{k,i}) \, ,
\end{equation}
or, in an even more compact form,
\begin{equation}\label{eq:alpha_compact}
    \vec{e} = H \vec{\alpha} \, ,
\end{equation} 
where we have introduced the $H$ matrix which is defined as
\begin{equation}
    H_{i,j} =  \frac{1}{n_i} \sum_k h_j(x_{k,i}) \, .
\end{equation}
Here, $H$ is a quadratic $M \times M$ matrix where $M$ is the total number of decay channels.
From the previous equations it is evident the role of the coefficients $\alpha_j$ which relate the NN classification response to the different decay-channel efficiencies.
The meaning of the $\alpha_j$ coefficients can be easily understood in the ideal case of a perfectly discriminating network.
In such a scenario $H$ takes the form of the identity matrix and $\alpha_j$ can be individually chosen to satisfy all possible combination of efficiencies, \emph{i.e.} we can choose $\alpha_{(j \vert j \in \mathcal{C})}=1$ for all control channels $\mathcal{C}$ in order to perfectly satisfy their efficiency constraints, and arbitrarily move $\alpha_{(j \vert j=s)}$ for the signal channel $s$ to get any possible values for the signal efficiency $e_s$.

In general, however,  the $H$ matrix will have non-diagonal terms that correlate the efficiency of the different decay channels.
The measurements carried out on the control channels, therefore, provide non trivial constraints on the signal efficiency.

\subsection{Optimisation procedure}

The goal of the algorithm is to find the solution that maximises possible shifts in the signal efficiency while maintaining the control measurements within their allowed range, as defined in Eqs.~\ref{eq:M_constr} and \ref{eq:eff_constr}.
This is achieved with an iterative procedure:
\begin{itemize}
    \item[i)] the NN is pretrained as a simple classifier, \emph{i.e.} minimising the cross-entropy loss between channels; 
    \item[ii)] for a given set of NN parameters $\theta$ the matrix $H$ is determined and the optimal values of the coefficients $\vec{\alpha}$ are calculated;
    \item[iii)] with the obtained values of $\vec{\alpha}$, the NN parameters are updated to improve the current solution;
\end{itemize}
with step ii) and iii) repeated for 1000 iterations. 
In each iteration the value of $e_s$ is recorded and the  solution that manifests the largest bias in the signal efficiency is returned at the end of the procedure. 
More details on the individual steps ii) and iii) are given in the next subsections.
In general, mismodelling of the efficiency can result in both over and underestimation of the total signal efficiency.
In the following, we focus on the minimisation of the signal efficiency; oppositely, in order to get the maximum allowed positive shift it will be sufficient 
to target $- e_s$ instead of $e_s$ in the minimisation process below.

\subsubsection{Determination of the coefficients \texorpdfstring{$\alpha_j$}{alphas}}\label{sec:LP}

In order to find the values of $\vec{\alpha}$ that allow the largest deviation in $e_s$ while satisfying the constraints from the control channels we have to solve the following system of linear equations
\begin{align}\label{eq:linprog}
   \begin{cases}
   \sum_j \alpha_j H_{i,j}   \in [V_{i}^{low};V_{i}^{high}]  & \forall i \in \mathcal{C} \, , \\
   e_s  = \sum_j \alpha_j H_{i,j} \to \min & \mathrm{for} \, i = s \, . \\
   \end{cases}
\end{align}
This linear programming (LP) problem can be solved numerically
and the minimisation of $e_s$ is performed with the SciPy python package with the use of the \texttt{scipy.optimize.linprog} function~\cite{2020SciPy-NMeth}.

\subsubsection{Update of the neural network}

The values of $\vec{\alpha}$ obtained in the previous step correspond to the optimal result for a fixed set of NN parameters, however, these can be modified to further improve the overall solution. 
The neural network weights $\theta$ are then updated based on the following loss function
\begin{equation}\label{eq:loss}
\ell(\theta) = \ell_s(\theta) + \ell_d(\theta) \, ,
\end{equation}
with
\begin{equation}\label{eq:loss_s}
\ell_s(\theta) = e_s
\end{equation}
driving the minimisation of $e_s$ with respect to the NN parameters $\theta$ and 
\begin{equation}\label{eq:loss_d}
\ell_d(\theta) = - \log \left\vert \det(H) \right\vert 
\end{equation}
is a regulariser term which is added to avoid $\det(H)=0$ and it keeps $H$ invertible during the optimisation process.
This term is found to help the stability of the process
since the determination of $\vec{\alpha}$ obtained in the previous step implicitly requires $H$ matrix inversion ($\vec{\alpha} = H^{-1} \vec{e}$) which can be numerically unstable.
The gradient of the loss $\ell(\theta)$ can then be calculated using the chain rule, \emph{i.e.} 
\begin{equation}
\frac{\partial \ell}{\partial\theta} = \sum_{j}\sum_{i}\frac{\partial H_{i,j}}{\partial\theta}\left(
\frac{\partial \ell_s(\theta)}{\partial H_{i,j}} +
\frac{\partial \ell_d(\theta)}{\partial H_{i,j}}
\right)
,
\end{equation}
with 
\begin{align}
\frac{\partial \ell_s(\theta)}{\partial H_{i,j}} &=
   \begin{cases}
   \alpha_j  & \mathrm{for} \, i = s \, , \\
   0 & \forall i  \in \mathcal{C} \, . \\
   \end{cases}
   \\
\frac{\partial \ell_d(\theta)}{\partial H_{i,j}} &= -\left[H^{-1}\right]_{j,i}
\label{eq:diff_ell_d}
\end{align}
where eq.~\ref{eq:diff_ell_d} follows from eq.~\ref{eq:loss_d} as shown in \cite{matrixcookbook}, while $\frac{\partial H_{i,j}}{\partial\theta}$ is computed with standard backpropagation techniques. 
Finally, the NN parameters $\theta$ are updated based on the overall loss function improvement $\theta \to \theta - \eta \frac{\partial \ell}{\partial\theta}$, where $\eta$ is the learning rate set to {0.0001}, and the previous step is repeated.
The overall optimisation procedure is schematised in Alg.~\ref{alg:training}.

\subsection{Smoothening of the output weighting function}\label{sec:grad_pen}

The solution obtained following the algorithm description presented in the previous section represents the largest possible variation of $e_s$ that keeps the control-channel measurements within their limits.
However, the resulting weighting function $w(x)$ can display extreme trends, such as fast-changing values, which can be mitigated by adding the following regulariser to the loss function
\begin{equation}\label{eq:gp}
\ell_g(\theta) = \frac1n\sum_k\left[\left(\frac{\left\Vert \nabla \vec{h}(x_k;\theta)\right\Vert}{p}-1\right)_+\,\right]^2 \, ,
\end{equation}
where $\left(\cdot\right)_+$ denotes a positive cut, \emph{i.e.} $\left(y\right)_+ = \max(y,0)$, 
$x_k$ is a dataset sampled randomly from the domain $\vec{x}$
and $p$ is a tunable parameter indicating the turn on of this gradient penalty (GP) term and is set to 0.5 in the rest of the paper.

\section{A concrete example: a branching fraction measurement}
\label{sec:toy_example}

In this section, we illustrate the use of the DL Advocate method with the example of a branching fraction measurement. Branching fractions are typically measured as a ratios of a given signal decay channel with respect to a normalisation channel
\begin{equation}\label{eq:RK}
    \mathcal{B}_{\rm sig}=\frac{N_{\rm sig}}{N_{\rm norm}}\cdot\frac{e_{\rm norm}}{e_{\rm sig}} \cdot \mathcal{B}_{\rm norm}
\end{equation}
where $\cal B$ is the branching fraction, $N$ is the observed yield and $e$ is the detector efficiency. 
It is evident from Eq.~\ref{eq:RK} that a problem in the efficiency estimation would unavoidably lead to an incorrect determination of the branching ratio.

For this example, we consider the branching fraction measurement of a hypothetical decay of the type $P\to V C$, where $V\to AB$ is a hypothetical intermediate resonance decaying into two given particles $A$ and $B$, and $C$ is a companion final-state particle. Branching ratios are typically normalised relative to  decays with similar topology, which we denote here with $P \to  X C$ channel, with $X \to AB$, where the branching ratios $P\to X C$ and  $X\to AB$ are assumed to be well known. For instance the Particle Data Group (PDG) lists several branching fractions for various systems, some of which have uncertainties as low as 3\%~\cite{PDG:2022pth}. In addition to the normalisation itself, crosschecks involving other control channels with well known branching fractions are often performed during experimental analyses. In the following, we consider the existence of a second control channel denoted with $P\to Y C$ with $Y\to AB$.

The advantage of the existence of a second control channel stands in the possibility to build ratio of branching fractions, \textit{i.e.} the same procedure developed for the signal is employed on the control channel by replacing in Eq.~\ref{eq:RK} the signal efficiency and observed yield with the corresponding control channel counterparts.
Ratios of branching fractions, in fact, are typically characterised by smaller uncertainties compared to single branching fractions, since common sources of uncertainties are cancelled out in the ratio.
For instance, the PDG reports ratios of branching fractions which are known with a precision at percent level~\cite{LHCb:2016ehk}.

In conclusion, in our toy example we assume the absolute branching ratio of the normalisation channel to be known with a precision of 3\%, the ratio of branching ratios between the two control channels to be known with a precision of 1\% and we train the DL Advocate to place an upper bound on the systematic associated to the measurement of signal branching ratio while obeying these constraints.

\subsection{Simulated dataset}\label{sec:simulation}

In order to explore this method in a concrete setting, 
we generate simulated events based on a typical measurement performed by the
LHCb experiment~\cite{LHCb:2008vvz}, as many heavy-hadron decays are measured by LHCb and it is therefore a natural testbed for our approach. 
In order to generate the representitive simulation we employ a fast simulation based on the the RapidSim package~\cite{Cowan:2016tnm}, which simulates decays of heavy hadrons with parameterised momentum smearing and representitive production kinematics for high energy collisions. Basic acceptance criteria are applied such as requiring each final state particle to be within $2< \eta < 5$, which is a crude estimate of the LHCb acceptance, and to have a minimum transverse momentum of $300~\mathrm{MeV}/c^2$. 
Finally, final state radiation effects are simulated with PHOTOS~\cite{Davidson:2010ew}, which play a minor role in the analysis.

We use the decay of a $B^+$ meson into a kaon and a pair of oppositly charged muons. This decay is chosen as a proxy for a generic three-body decay as several control channels exist such as decays involving the $\jpsi$ and $\psitwos$ mesons~\cite{LHCb:2014cxe} which make it a plausible test case for the method. As there is no detailed particle identification response in our analysis, we use this decay as a representation of the general three-body $P \to ABC$ particle decay introduced above, where the two muons, represented here as $A$ and $B$, originate from an intermediate resonance of variable mass, $m_V$, and the kaon represents particle $C$. 
One hundred thousand events are generated for each considered value of $m_V$.

While the generated samples do not take into account reconstruction effects specific to the LHCb detector, they provide an excellent template to demonstrate the proposed method on a simplified example.

\subsection{Analysis setup}\label{sec:setup}

\begin{figure}[t]
\centering
\includegraphics[width=0.7\textwidth]{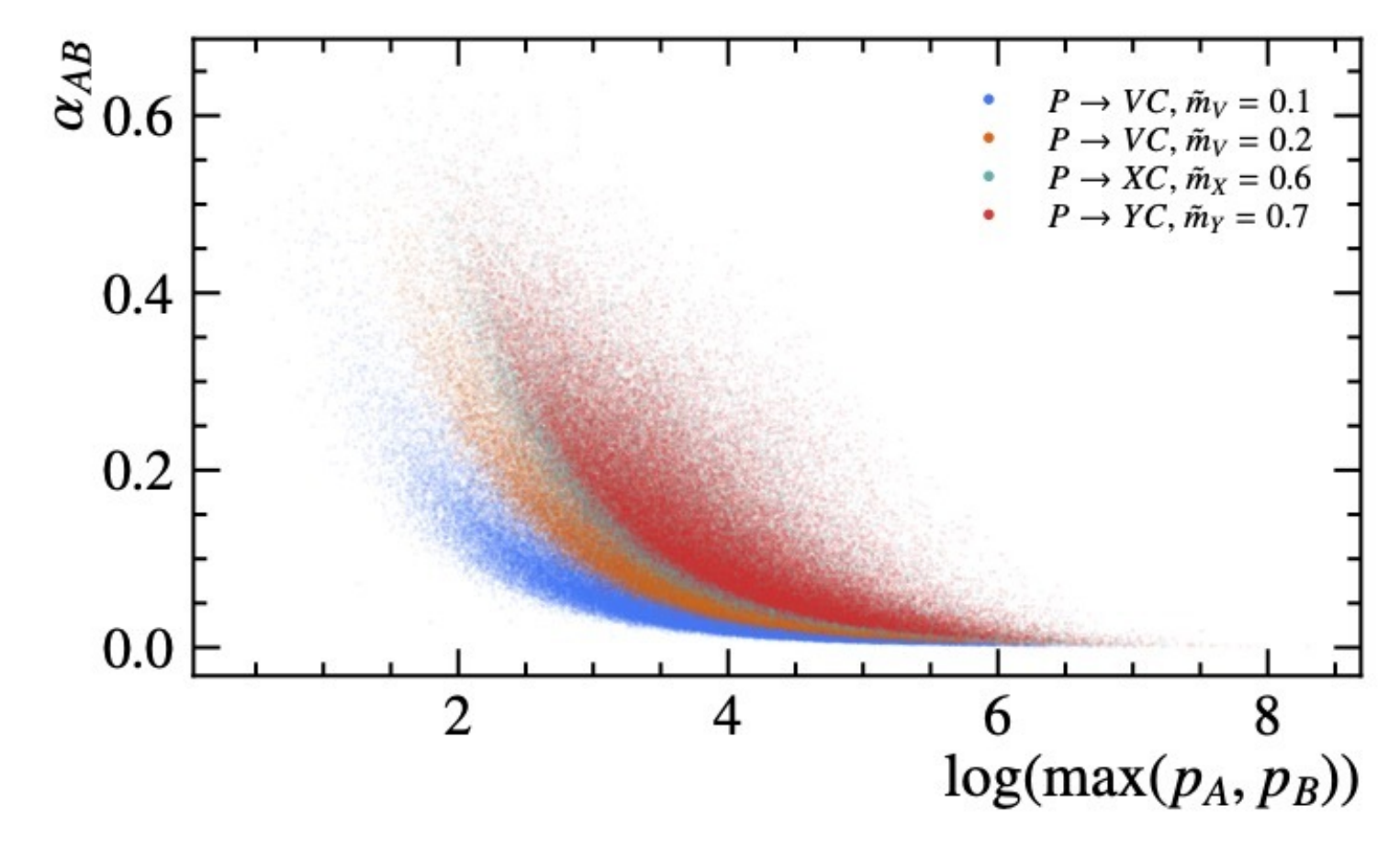}
\caption{Distribution of events in the kinematic plane $\max( p_A, p_B), \, \alpha_{AB}$ for two choices of signal mass $\tilde{m}_V = 0.1$ and 0.2, as well as for the control channels.}
\label{fig:scatter}
\end{figure}

In order to determine how a potential mismodelling of the efficiency could affect the branching fraction measurement of $P\to V C$ decays while keeping unchanged the value of the crosscheck provided by the control channel $P \to Y C$, we need to consider all possible differences among the considered decay channels. Due to the different resonance masses $m_R$, with $R \in \{V, X, Y\}$, signal and control modes are characterised by different kinematic distributions. In the following we use the normalised resonance mass $\tilde{m}_R=m_R/m_P$ as a more generalisable kinematic variable. This resonance mass $\tilde{m}_R$ can be expressed as $\tilde{m}_R^2 = 2 p_A p_B (1 - \cos \alpha_{AB})/m_P^2$, \emph{i.e.} the mass of the resonance is unequivocally determined by the momentum of the two final state particles $AB$ and their opening angle. 
As an example, we take $X$ and $Y$ particles to have similar normalised masses, \emph{i.e.} $\tilde{m}_{X} \simeq 0.6$ and $\tilde{m}_{Y} \simeq 0.7$, which results in similar kinematic distributions, while $\tilde{m}_V$ is allowed to span all the kinematically allowed range, with the result that the further $\tilde{m}_V$ is from $\tilde{m}_{X, Y}$ the bigger the difference in their kinematic distributions will be (see Fig.~\ref{fig:scatter}).

For the example illustrated in this paper, we will therefore focus on the study of the momentum, transverse momentum and opening angle, or combination of those. 
In high-energy-physics experiments, the efficiency to reconstruct and select a given particle is typically dependent on its (transverse) momentum, which makes particularly important --- and potentially prone to hidden systematics --- to have good control of the detector efficiency as function of those variables. 

In the following, we will therefore train the DL Advocate to evaluate what is the maximum impact that a hypothetical mismodelling of the efficiency can have on the determination of the $P\to V C$ branching ratio under different values of $\tilde{m}_{V}$.
It is also important to stress that the goal of the algorithm shall not be to reconstruct $\tilde{m}_V$, which is what occurs if the two momenta and the opening angle of the $AB$ particles are given to the network, but rather to explore possible undetected patterns in the efficiency response of the detector which may lie hidden in a subset of such kinematic variables.

Three different sets of input variables are therefore considered for this study:
\begin{itemize}
    \item $ x = \{ p_T^{A}, \, p_T^{B} \}$, the two transverse momenta of particles A and B;
    \item $ x = \{ \max p, \, \alpha_{AB} \}$, the maximum momentum of the two particles A and B and their opening angle;
    \item $ x = \{ \max p_T, \, \alpha_{AB} \}$,
    the maximum transverse momentum of the two particles A and B and their opening angle;
\end{itemize}
The DL Advocate algorithm is then trained following the methodology described in Sec.~\ref{sec:method} with the constraints imposed by the normalisation and control channels taken into account as discussed in Sec.~\ref{sec:LP}.
In particular, the constraints takes the form
\begin{eqnarray}
    & & \mathcal{B}(P \to X C) \propto  e_{P \to X C} \in [-3 \%, 3\%] \,  , \\
    & & \frac{\mathcal{B}(P \to Y C)}{\mathcal{B}(P \to X C)}  \propto \frac{e_{B \to Y C}}{e_{P \to X C}} \in [-1 \%, 1\%] \, .
\end{eqnarray}
where $e_i$ should be interpreted as relative variations of the efficiency with respect to the hypothesis of perfectly modelled efficiency (as defined in Eq.~\ref{eq:eff_def}) and the former equation assumes perfect knowledge of the production of the parent particle $P$.
Finally, in order to evaluate the dependence of the maximum allowed bias on $\tilde{m}_V$, the training is repeated for different values of the normalised resonant mass ranging from 0.1 to  0.9.

\subsection{Training and results}
\label{sec:results}

\begin{figure}[t]
\centering
\includegraphics[width=0.85\textwidth]{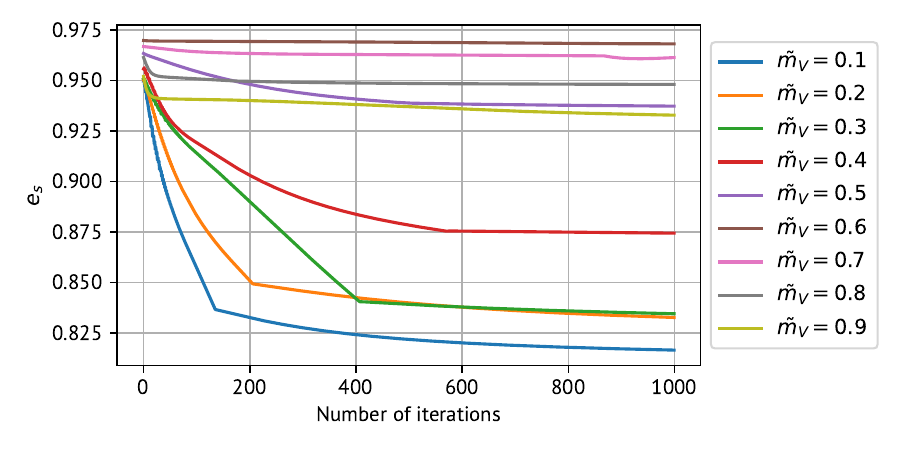}
    \caption{Evolution of the signal efficiency $e_s$ for the different values of $\tilde{m}_V$ when trained with the set of features $ x = \{ p_T^{A}, \, p_T^{B} \}$.}
    \label{fig:training_loss}
\end{figure}

Figure~\ref{fig:training_loss} shows the evolution of $e_s$ during the training process for the different values of $\tilde{m}_V$ and for the selected input features $ x = \{ p_T^{A}, \, p_T^{B} \}$. 
A good learning curve is observed for all cases, with an almost-optimal solution obtained after less than 500 iterations.
A very similar pattern is also seen for the other pairs of features used.

\begin{figure*}[t]
\centering
\includegraphics{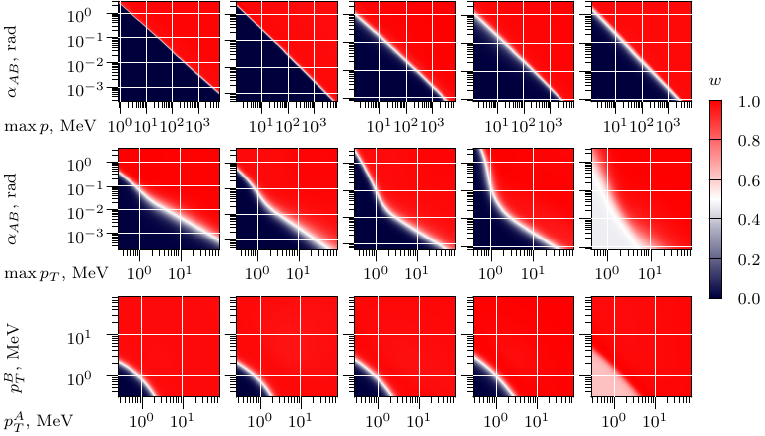}
\caption{Resulting weighting maps describing the efficiency mismodelling as function of the input variables for the three sets of considered features $\{\max p, \alpha_{AB}\}$ (top), $\{\max p_T, \alpha_{AB}\}$ (middle) and $\{ p_T^{A}, p_T^{B}\}$ (bottom)
for $\tilde{m}_V = \{0.1, 0.2, 0.3, 0.4, 0.5\}$ (from left to right). Values of $\tilde{m}_V \geq 0.6$ are not shown since no significant deviations from unity are visible in the entire feature plane.}
\label{fig:NN_weights}
\end{figure*}

\begin{figure*}[h!]
\centering
\includegraphics[width=0.48\textwidth]{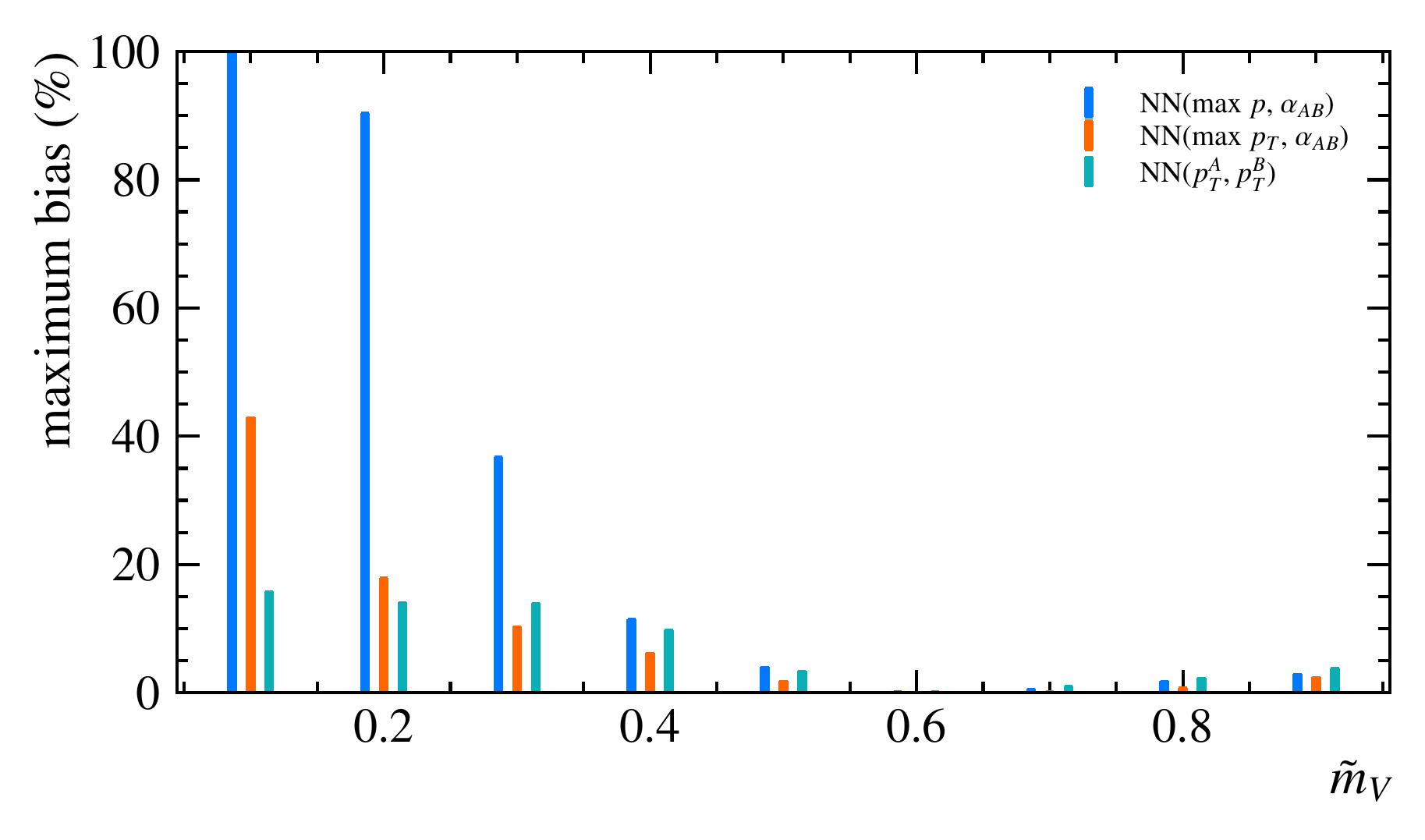}
\hspace{1mm}
\includegraphics[width=0.48\textwidth]{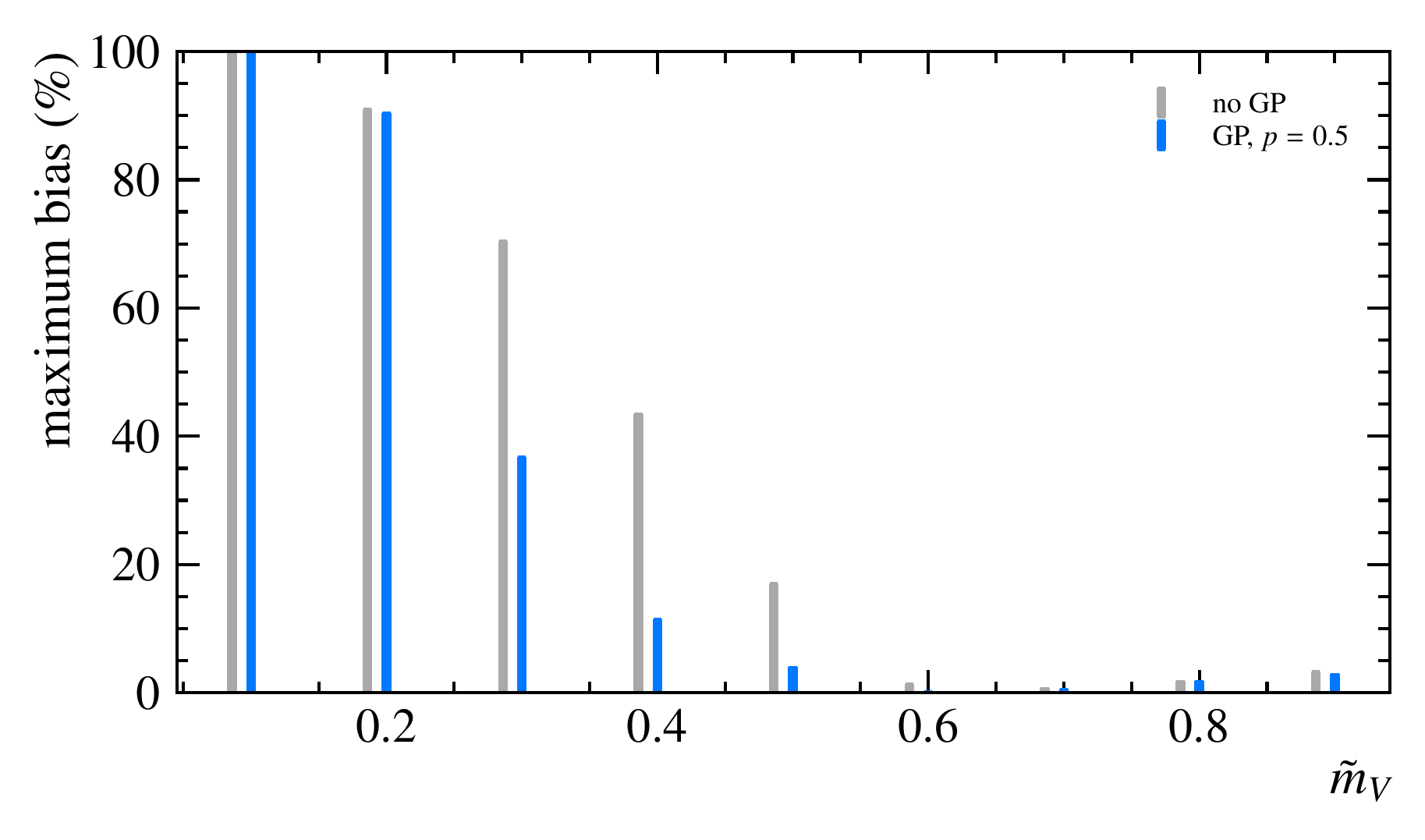}
\caption{Left: maximum allowed bias obtained on $P \to V C$ branching ratio by running the DL Advocate algorithm with the three studied sets of features at different masses. All cases have been trained with a gradient penalty term set to 0.5.
Right: maximum allowed bias obtained using $\{\max p, \alpha_{AB}\}$ as input features with and without the gradient penalty term introduced in Eq.~\ref{eq:gp}.
}
\label{fig:bias}
\end{figure*}

Figure~\ref{fig:NN_weights} shows the resulting weighting functions obtained from the training of the network for the three considered sets of variables and for different values of $\tilde{m}_V$.
For cases where $\tilde{m}_V < \tilde{m}_{X}$, the obtained solution shows a clear split of the 2D plane, with a large fraction of signal events that receive a weight close to zero.
This is due to the larger separation in the kinematic space between signal and control channels, which allows to find a solution that strongly affects the former while keeping unchanged the other two.
On the other hand, for signal masses between or above the $X$ and $Y$ particles, the strong overlap in the kinematic distributions leaves very little to no room for efficiency mismodelling to be able to modify the signal branching ratio while satisfying the constraints from the control channels.

We can now quantitatively define the observed bias in each decay mode as the deviation of its integrated efficiency with respect to the hypothesis of perfect efficiency modelling, \textit{i.e.} $(1-e_i)$.
The maximum allowed shift on the determination of the signal efficiency, and hence the signal branching ratio, obtained in the different tested configurations is illustrated in Fig.~\ref{fig:bias}.
We can draw the following conclusions:
\begin{itemize}
    \item as expected, the lower the normalised signal mass $\tilde{m}_V$ is, the bigger the bias is allowed to be; this is due to the large separation in the feature space between signal and control channels, which allows a mismodelling of the efficiency that only affects the signal decay channel; 
    \item vice versa, it is nearly impossible for signal with mass similar to the ones of the normalisation/control channels to suffer from uncontrolled systematic effects;
    \item in all cases, the set of variables that allows the largest bias is given by the combination  $\{ \max p, \, \alpha_{AB} \}$, which is natural being the one with the largest correlation with the resonant mass;
    \item as expected, the gradient penalty term applied to the loss function as described in Sec.~\ref{sec:grad_pen} reduces the overall allowed systematic effect.
\end{itemize}
We note that, in most of the cases the obtained solution presents region of the parameter space where the mismodelling of the efficiency is maximal (\emph{i.e.} an efficiency very close to zero and/or infinity).
In addition, despite the gradient penalty term added to the loss function as in Eq.~\ref{eq:gp},  rather steep transitions between well-modelled and badly-modelled regions are still visible in Fig.~\ref{fig:NN_weights}.
Nevertheless, the effectiveness of this term is clearly visible from the comparison of the solution obtained with and without it as shown in Fig.~\ref{fig:bias} (right) and a further tuning of this parameter goes beyond the scope of this paper.

\subsection{Implications on differential control-channel measurements}\label{sec:dataMCagreement}

\begin{figure*}[t]
\centering
\includegraphics[width=0.45\textwidth]{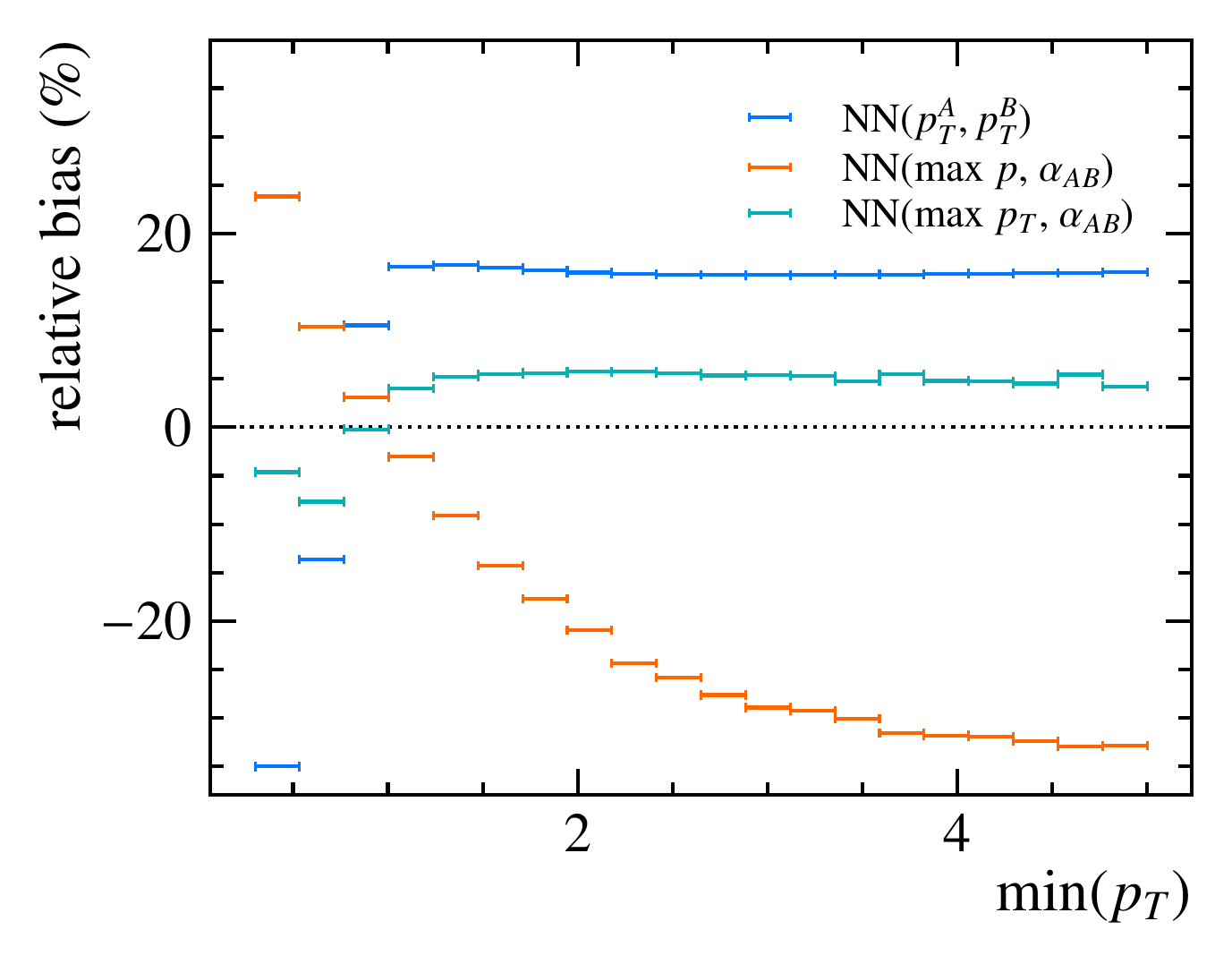}
\hspace{2mm}
\includegraphics[width=0.45\textwidth]{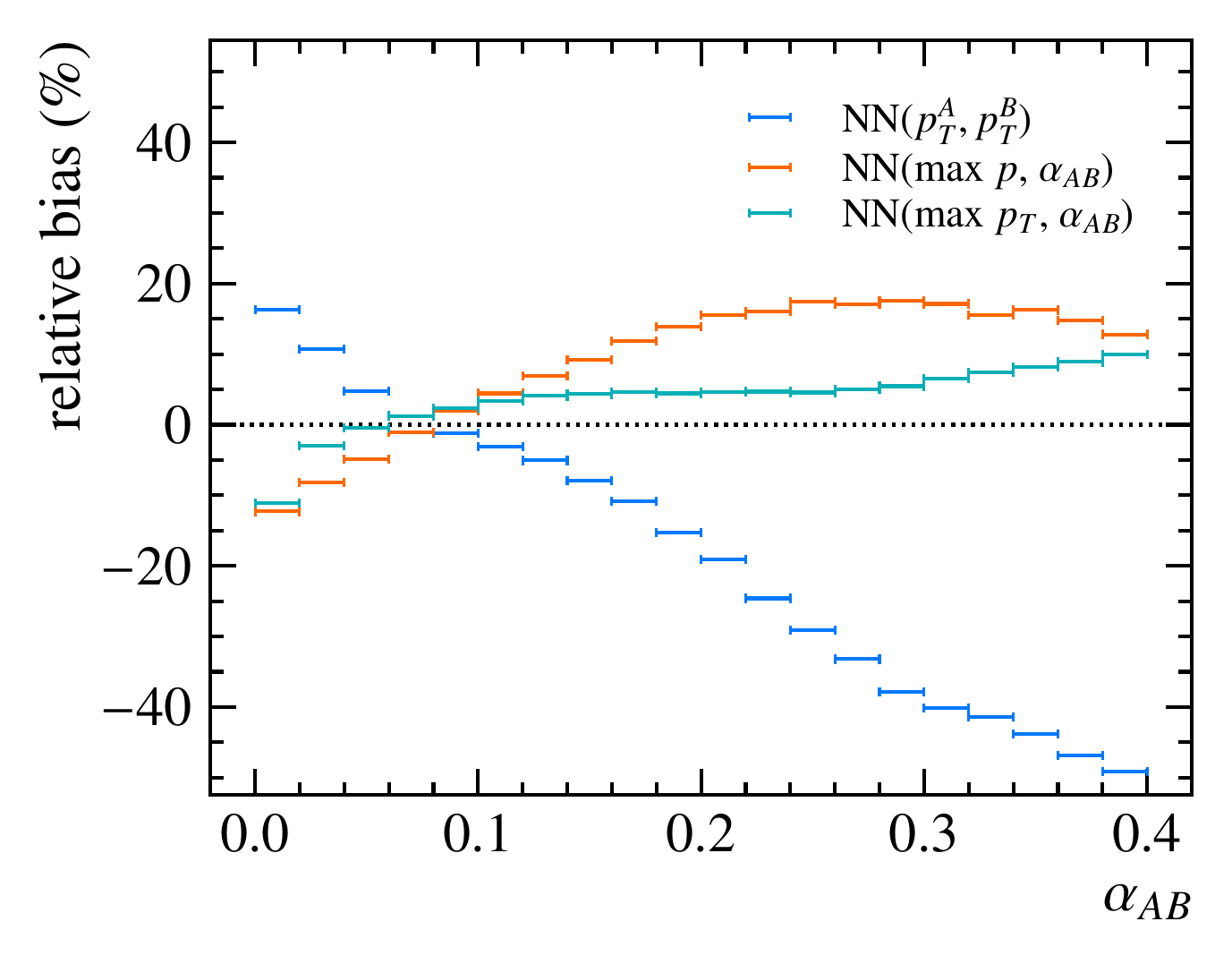}
\caption{Expected relative bias on differential branching fraction measurement for  the $P \to X C$ control channel as function of the minimum transverse momentum (left) and opening angle (right). 
The algorithm has been trained with a signal of $\tilde{m}_V = 0.3$, while the result for the three tested configurations is shown with different colours.}
\label{fig:dataMCagreement}
\end{figure*}

In the toy study presented above, we limited the role of the control channels to the sole contraint provided by their integrated branching fractions.
In real analyses, however, the use of simulated datasets undergoes a rigorous validation process, which once again makes use of well-known and largely produced  calibration channels.
One of the  additional crosschecks which can be naturally performed on the control channels consists in looking for possible biases as function of the relevant kinematic variables. 
A systematic uncertainty can in fact escape the bound imposed by the total branching fraction measurement while appearing in its differential distribution.
This is exactly the case studied in the previous section.
Figure~\ref{fig:dataMCagreement} shows the bias expected for the different obtained solutions as function of the minimum transverse momentum and opening angle of the considered control channel.
The strong visible trend demonstrates that, assuming a sufficient statistical precision, the systematic uncertainty  postulated by the DL Advocate algorithm in the previous section would have been revealed by analysing the consistency of the control-channel differential branching fraction measurement.
For example, in Ref.~\cite{LHCb:2014cxe}, which measures the branching fraction of 
$B\to K^{(*)}\mumu$ decays, the biases shown in Figs.~\ref{fig:bias} and \ref{fig:dataMCagreement} would have been detected by the experimental crosschecks performed in the analysis thanks to the large datasets  collected for the $B\to J/\psi K^{(*)}$ normalisation channels, which are of the order of hundreds of thousands of events.

One of the limitation of the standard analysis procedure, which consists in manually validating the simulation by looking at different relevant variables, is that it is limited to the statistical power and kinematic distribution of the control channels.
Therefore, a systematic effect may still escape this validation procedure if it is limited to a region poorly populated by the control samples or it appears with complex correlations among different or unexpected sets of variables.
On the opposite, the DL Advocate algorithm presented in the previous section can  be extended to include any arbitrary set of constraints, such as the differential control channel measurements discussed above, and provide a solution that would pass all the standard crosschecks of this kind of analyses.
A complete implementation of all the available constraints goes beyond the scope of this paper, however, we can expect them to play a significant role in reducing the room for possible undetected hidden systematic uncertainties.

Finally, beyond all internal crosschecks that can be developed within a given analysis, one of the strength of the proposed method is that one can systematically explore all the repercussions that a solution found by the DL Advocate may have on other observables and/or decay channels.
For example,  one can check the impact that a certain solution obtained   investigating  systematic uncertainties on the signal branching fraction has on the mismodelling of the angular distributions of the same decay or, vice versa, how measuring the angular distribution of the decay can further constrain the impact of hidden systematic uncertainties on the branching fraction determination.

\section{A Reinforcement Learning approach}\label{sec:RL}

As already mentioned, the method proposed in this paper can only reach its full potential when applied to \emph{low-level} detector information.
The use of detector-related quantities, such as hits positions, energy clusters, magnetic fields, etc., introduces several additional complications to the problem, since it requires an accurate description of the detector and, most importantly, an iterative refinement of the simulation, \textit{i.e.} the  interaction of the particles with the detector has to be re-evaluated for every considered modification.
On the other hand, the inclusion of low-level quantities would 
provide a general tool that is applicable to any measurement that relies on simulation, enabling a systematic evaluation of all possible mismodellings of the detector response.
While a complete solution to this problem is out of the scope of this paper, 
we illustrate with a simplified example the use of Reinforcement Learning (RL) to potentially tackle this difficult task.

Reinforcement learning~\cite{sutton2018reinforcement,li2018deep} recently achieved impressive results in many domains of applied research, such as robotics~\cite{kalashnikov2018qtopt}, self-driving cars~\cite{kiran2021deep}, gaming~\cite{silver2017mastering}.
Coming to high energy physics, it has been mainly suggested for jets reconstruction~\cite{Carrazza:2019efs,Cranmer:2021gdt} and on-line control system for accelerator machines~\cite{StJohn:2020bpk}.
In this section we discuss the possibility to train a RL \emph{agent} to play the role of the devil’s advocate (RL Advocate).
The goal of the algorithm is unchanged, \emph{i.e.} trying to find possible mismodelling  of the efficiencies that can affect a certain set of observed measurements; however,  we need to introduce some new concepts in order to formalise the problem within a RL approach.

\begin{figure}[t]
\centering
\includegraphics{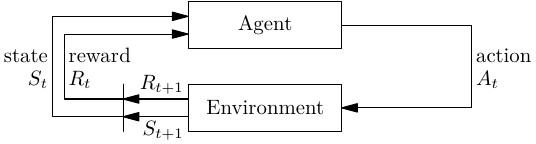}
\caption{Schematic representation of agent-environment interaction~\cite{sutton2018reinforcement}.}
\label{fig:RL_diag}
\end{figure}

Reinforcement learning algorithms are designed to train an \emph{agent} via continuous interaction with an external environment.
At each time step $t$, the agent makes an observation of the environment's \emph{state}, $S_t$, and, based on such observation, it undertakes a certain \emph{action} $A_t$.
In turn, as a consequence of this action, the agent will find itself in a new state $S_{t+1}$, while receiving from the environment a numerical reward $R_{t+1}$.
Figure~\ref{fig:RL_diag} illustrates the described process.
Finally, the decision-making of the agent, also called \emph{policy}, $\pi$, is trained to maximise the expected total reward over the long run.

\subsection{Environment's setup}

Identically to the classifier-based approach discussed in Sec.~\ref{sec:method}, also in this case we intend to describe
possible mismodelling of the efficiency by introducing a per-event weighting function 
$w(x)$ which can depend on a certain set of input features $x$.
However, instead of training a neural network to determine the mismodelling map,
we define a parametric expression of the weighting function itself,
$w(\tilde{\eta},x)$, where $\tilde{\eta}$ is the set of parameters that describes the detector response which needs to be determined.
This approach has some advantages, \emph{e.g.} we avoid the risk of
having sharp (unphysical) changes of efficiencies, as well as disadvantages, 
\emph{e.g.} the requirement of a parametric expression in the formulation of the problem, first, looses generality compared to the NN output (given the enormous number of parameters that characterises a NN allowing it to be able to approximate  any function in its domain), and second, it definitely requires some physics intuition.

The goal of the RL agent can therefore be formulated as finding the values of the parameters $\tilde{\eta}$
that best satisfies the agreement with the experimental measurements.
We can then define:
\begin{itemize}
    \item the \emph{state}, $s \equiv \tilde{\eta}$ (renamed for convenience), the list of the parameters used to describe the weighting function;
    \item the possible \emph{actions}, which consist of increasing or decreasing each of the $\tilde{\eta}$ parameter by a discrete quantity;
    \item the \emph{reward} system,
    \begin{align}
    r = 0.01 \times
    \begin{cases}
    - \chi^2 / N_{\mathrm{meas}}  &  \mathrm{if} \; \chi^2 / N_{\mathrm{meas}} > 3, \\
    - \chi^2 / N_{\mathrm{meas}} + 10 \, (3 - \chi^2 / N_{\mathrm{meas}})   &  \mathrm{if} \;  \chi^2 / N_{\mathrm{meas}} \in [0.1,\, 3], \\
    10^3  &  \mathrm{if} \; \chi^2 / N_{\mathrm{meas}} < 0.1, \\
    \end{cases}
    \label{eq:reward}
    \end{align}
    with $N_{\mathrm{meas}}$ the number of considered measurements and
    $\chi^2 \equiv \sum_{i=1}^{N_{\mathrm{meas}}} \left( \frac{M_i - \mu_i}{\sigma_i} \right)^2$
    where, recalling the notation introduced in Eq.~\ref{eq:syst_eq_compact},  $M_i = \mathcal{F}(\tilde{\eta})$ are the values of the different measurements evaluated at each time step based on the evolving values of the $\tilde{\eta}$ parameters, 
    while $\mu_i$ and $\sigma_i$ are the corresponding central values and uncertainties  measured by the experiment.
\end{itemize}
The scheme of Eq.~\ref{eq:reward} is designed to return a negative reward (proportional to $\chi^2$)
when the agent is far from finding a good solution that accommodates all the considered measurements (large $\chi^2$ values).
On the other hand, the agent is \emph{encouraged} with an increasingly positive reward once it reaches the condition $\chi^2 / N_{\mathrm{meas}} < 3$.
Finally, in the event the agent finds a good compatibility between the obtained solution and the experimental measurements, which is  defined in Eq.~\ref{eq:reward} as $\chi^2 / N_{\mathrm{meas}} \leq 0.1$, a large positive reward is assigned and the episode is terminated.
Ultimately, since $\chi^2$ can take on very large values,
a scaling factor of 0.01 is added to improve the convergence of the algorithm.

Each episode is run for a maximum of 200 time steps, while the 
 size of each parameter's update is adjusted to decrease 
from 0.1 to 0.01,
according to the returned $\chi^2 / N_{\mathrm{meas}}$ value,
\emph{i.e.} the closer the algorithm is to the best solution 
the smaller the parameter's update will be.

We implemented the described environment within the
RLlib python library~\cite{liang2018rllib},
which offers a great variety of Reinforcement Learning algorithms
ready for use.
In the following studies, we will employ
deep $Q$-networks (DQN)~\cite{DQN,DoubleDQN,DuelingDQN}, 
which are designed to learn the action-value function,
or $Q$-value, defined as 
\begin{equation}
Q(s,a) =  \mathbb{E}_\pi \left[ \sum_{k=0}^\infty \gamma^k R_{t+k+1} \bigg\vert S_t=s, A_t=a \right] 
\end{equation}
where $\gamma$ is a discount factor that weights future rewards compared to present ones, whose default value is set to 0.99.
In general, the $Q$-value quantifies \emph{how good} is to take action $a$ while being in state $s$; learning the $Q$-value function is the target of the DQN training procedure.
Once a good approximation of the $Q$-values is achieved,
it will be sufficient to always select the action with the highest 
$Q$-value, namely following a \emph{greedy} policy, in order to 
find the optimal solution.
However, better training performances are typically achieved by following
the so-called $\varepsilon$-\emph{greedy} policy, where a random action is selected a fraction $\varepsilon$ of the times. 
This behaviour allows more exploration for new --- and potentially better --- actions and it is found to help the convergency of the training process.
In the following, the value of $\varepsilon$ is chosen to decrease exponentially from 1.0 to 0.1 with a decay length of $10^4$ steps.

\subsection{RL Advocate applied to \texorpdfstring{$P_{5}^{'}$}{P5'}}

This section illustrates the use of Reinforcement Learning to play the devil's advocate with a second concrete example inspired by a set of measurements that shows some tension with respect to the Standard Model prediction.
In principle, the same study could be conducted with the linear programming method presented in Sec.~\ref{sec:method} (after an appropriate modification of the loss definition) as well as RL be applied to the branching fraction example of Sec.~\ref{sec:toy_example}. 
In both cases similar results and conclusions have to be expected given the simple nature of the examples under study and agreement between the simple optimisation and RL results shown later.
In the following, we present the efficacy of the RL Advocate applied on a physics example of high interest for the community, \emph{i.e.} the $P_{5}^{'}$ anomaly~\cite{Altmannshofer:2013foa}.

\subsubsection{The \texorpdfstring{$P_{5}^{'}$}{P5'} measurement}

Rare decays proceeding via $b \to s \ell \ell$ transitions are a sensitive probe of physics Beyond the Standard Model.
In particular, potentially yet-undiscovered particles may contribute to the decay process and modify, among other properties, the angular distribution of the final state 
particles~\cite{Gauld:2013qja,Buras:2013qja,Altmannshofer:2013foa,Crivellin:2015era}.
Precision measurements of these angular observables are therefore a fundamental test of the Standard Model.

In recent years, different experiments~\cite{LHCb:2020lmf,Wehle:2016yoi,ATLAS:2018gqc,CMS:2017rzx} measured a deviation in one of the angular observables of $B^0 \to K^{*0} \mumu$ decays, namely $P_{5}^{'}$, in a region of $q^2$ between 4 and 8~GeV$^2$, where $q^2$ is defined as the di-muon invariant mass squared.
In the following, we will use the example of the LHCb measurement of $P_{5}^{'}$ in the $[4.0, \, 6.0]$~GeV$^2$ $q^2$ bin~\cite{LHCb:2020lmf} 
\begin{equation}\label{eq:s5_lhcb}
    P_{5}^{'} = - 0.439 \pm 0.117
\end{equation}
which is found to be higher than the Standard Model value by a factor of 30-40\%, depending on the considered theoretical predictions~\cite{Bharucha:2015bzk,Altmannshofer:2014rta,Khodjamirian:2010vf,Descotes-Genon:2014uoa}.

In the following, we will run the RL Advocate to test whether an uncontrolled modification of the efficiency can cause a shift in the $P_{5}^{'}$ observable capable of explaining the observed discrepancy.

\subsubsection{Determination of \texorpdfstring{$P_{5}^{'}$}{P5'} and choice of the weighing function}

Three decay angles are necessary to describe the angular distributions of $B^0 \to K^{ *0} \mumu$ decays, namely
$\vec{\Omega} = \{ \theta_K, \, \theta_\ell, \, \phi \}$, defined as in Ref.~\cite{LHCb:2015svh}.
Starting from this angular definition, one can define  a  basis of angular functions, whose coefficients --
the angular observables -- depend on the underlying physics model~\cite{Altmannshofer:2008dz}.
The $P_{5}^\prime$ observable is  defined as a combination of the angular coefficient $S_5$ and the fraction of longitudinal polarisation of the $K^{*0}$ meson $F_L$, \emph{i.e.}
$P_5^\prime = \frac{S_5}{\sqrt{F_L(1-F_L)}}$~\cite{DescotesGenon:2012zf,Matias:2012xw}.

In the case of a pure signal sample, $F_L, \, S_5$ and consequently $P_{5}^{'}$ can be easily extracted using the \emph{method of moments}~\cite{LHCb:2015svh}, 
\emph{i.e.} averaging the value of the relevant angular functions $f_j$ over the collected sample, 
which gives $F_L = 2-\frac{5}{2}M_{1s}$ and $S_{5} = \frac{5}{2} M_5$, where the relevant angular moments are defined as
\begin{equation}
M_{1s(5)} = \frac{1}{N} \sum  f_{1s(5)}(\vec{\Omega}_i) ,
\end{equation}
with $f_{1s} = \sin 2 \theta_K$ and $f_5 = \sin 2 \theta_K \sin \theta_\ell \cos \phi$.

The loss of efficiency due to imperfect experimental detection can be taken into account by introducing a series of weights and modifying the 
determination of the moment $M_j$ to
\begin{equation}
M_{1s(5)} = \frac{1}{\sum w_i} \sum w_i f_{1s(5)}(\vec{\Omega}_i) ,
\end{equation}
where $w_i$ represents a per-event weighting function.
In general the weight $w_i$ encodes the distorsion occurred to the signal angular distributions due to the finite detector efficiency, which is studied with high accuracy by the different experiments~\cite{LHCb:2020lmf,Wehle:2016yoi,ATLAS:2018gqc,CMS:2017rzx}.
In the following, however, we will uncouple all known detector effects, \emph{i.e.} as in the previous sections we assume perfect estimated efficiency, and we entirely identify 
the weight $w_i$  with the mismodelling of the efficiency under investigation.
Finally, for this study we employ a fast simulation of one million events of $B^0 \to K^{ *0} \mumu$ decays, with $K^{*0} \to K^+ \pi^-$, obtained with the RapidSim package~\cite{Cowan:2016tnm}
run under the same conditions described in Sec.~\ref{sec:simulation}.

Before entering the core of the problem, it is interesting to test whether the solution obtained in the previous section
can have any sort of impact on the angular observables $P_{5}^{'}$.
We note that none of the solutions found in Sec.\ref{sec:results} can cause a shift in $P_{5}^{'}$, this is because $P_{5}^{'}$ (and the angular observables in general) 
does not depend on the total signal efficiency, as it was for the branching fraction, but it can only be affected by variations of the efficiency in the angular space.
Therefore, in order to create a shift in $P_{5}^{'}$, 
we need a  mismodelling of the efficiency that goes beyond the simple dependency on the total kinematic of the event.
One possibility would be to introduce a detector response that is different for the two muons.
In particular, one can consider a muon efficiency that depends on the relative charge between the muon and the kaon.
Such scenario can in principle occur if there are hidden correlations between the hadronic system and the muons in the reconstruction of the decay.
In the ideal case, one may want the \emph{agent} to reach this conclusion by itself and autonomously take the initiative to split the efficiency weighting function by muon charge. 
While this possibility may appear exciting, it introduces an enormous amount of complications in the definition of the \emph{state--action} space,
\emph{e.g.} a variable number of available actions and/or variable dimensionality of the state, and it is left to future work.

In the following, we will therefore  program the \emph{agent} to have a  mismodelling function which is different for the two muons 
and we define the total per-event weight
\begin{equation}
w_i(x) = w^+ \cdot w^-   ,
\end{equation}
where $w^+$ is the mismodelling function associated to the muon with the same charge of the kaon, \emph{e.g.} $\mu^{+(-)}$ for a $K^{+(-)}$ in the final state,
while $w^-$ the one associated to the oppositely charged muon, \emph{e.g.} $\mu^{-(+)}$ for a $K^{+(-)}$ in the final state.

Concerning the choice of the functional form for $w^+(w^-)$, we limit the study reported in this example to a simple linear dependence on the muon transverse momentum, \emph{i.e.}
\begin{eqnarray}
w^+(p_T) = (1 - k_+ ) + 2 k_+ \cdot p_T^{(\mu^{SS})} \, ,\\
w^-(p_T) = (1 - k_- ) + 2 k_- \cdot  p_T^{(\mu^{OS})} \, ,
\end{eqnarray}
where $k_+$ and $k_-$ are the two parameters to be determined by the RL Advocate and $\mu^{SS}(\mu^{OS})$ indicates the same-sign (opposite-sign) muon with respect to the kaon. 
In addition, in order to avoid unphysical negative values of the efficiency, a lower bound of 0.1 is applied to the weights $w^+$ and $w^-$, 
\emph{i.e.}  $w^\pm = \max(0.1, w^\pm)$.
In principle the mis-modelling function can be of any form. However, we assume this simple parameterisation as a proof-of-principle of the proposed method for ease of comparison with simple optimisation for validation purposes. 

\subsubsection{Training and results}

We train the RL Advocate under the configuration described above with two different targets
\begin{itemize}
    \item[(I)] First, we only focus on signal $B^0 \to K^{*0} \mumu$ decays and try to see if we can find a mismodelling of the efficiency that could describe a shift in $P_{5}^{'}$ of the order of what it is seen in the experiment; 
    \item[(II)] second, we introduce the control channel $B^0 \to \jpsi K^{*0}$, with $\jpsi \to \mumu$, and check if a systematic bias in the determination of $P_{5}^{'}$ is still possible once we impose the constraints derived from the precise measurement of the angular observables in this control channel.
    Similarly to the previous example, in fact, the resonant mode provides an important validation of the efficiency estimation.    
\end{itemize}
In concrete terms, the difference between the two training configurations stands in the composition of the \emph{reward} system defined in Eq.~\ref{eq:reward}.
In the first case, only one measurement is considered and the $\chi^2$ entering in the reward is simply
\begin{equation}\label{eq:chi2}
    \chi^2_{\mathrm{case-I}} =  \bigg( \frac{P_{5}^{'}(k^+, k^-)  - {P_{5}^{'}}^{\mathit{meas}}}{\sigma^{\mathit{meas}}_{P_{5}^{'}}} \bigg)^2 \, ,
\end{equation}
where ${P_{5}^{'}}^{\mathit{meas}}$ and $\sigma^{\mathit{meas}}_{P_{5}^{'}}$ are the central value and uncertainty measured by the experiment, while $P_{5}^{'}(k^+, k^-)$ is the modified value of $P_{5}^{'}$ based on the variation of the mismodelling parameters $k^+, k^-$.
On the other hand, in order to include the constraint derived from the control channel, it is sufficient to add to the $\chi^2$ of Eq.~\ref{eq:chi2} the term corresponding to the control channel measurement
\begin{equation}
    \chi^2_{\mathrm{case-II}} =  \bigg( \frac{P_{5}^{'}(k^+, k^-)  - {P_{5}^{'}}^{\mathit{meas}}}{\sigma^{\mathit{meas}}_{P_{5}^{'}}} \bigg)^2 + \bigg( \frac{{P_{5}^{'}}^{\jpsi}(k^+, k^-)  - {P_{5}^{'}}^{\mathit{SM},\jpsi}}{\sigma_{{P_{5}^{'}}^\jpsi}} \bigg)^2  \, ,
\end{equation}
where ${P_{5}^{'}}^{\mathit{SM},\jpsi}=0$ is Standard Model value of $P_{5}^{'}$ for $B^0 \to \jpsi K^{*0}$ decays, $\sigma_{{P_{5}^{'}}^\jpsi}$ is assumed to be 0.0015, which is the typical uncertainty associated to the control channel measurement~\cite{LHCb:2014xzf,Belle:2014nuw} and ${P_{5}^{'}}^{\jpsi}(k^+, k^-)$ is the alteration of the control channel value of $P_{5}^{'}$ obtained by running the RL Advocate.

Table~\ref{tab:result_RL} shows the results obtained by running the RL Advocate for 3000 episodes. 
We note that, in this simplified example, finding the best solution for $\{ k_+, \, k_-\}$ is equivalent to find the minimum of the $\chi^2$. 
Therefore, we can validate the result obtained by the trained RL agent with the one derived by a simple $\chi^2$ minimisation, also shown in Tab.~\ref{tab:result_RL} for completeness.
We can draw the following conclusions
\begin{itemize}
    \item in case-I, the solution $\{ k^+=0.04, k^-=-1.09\}$ found by the RL Advocate can result in a positive shift in the value of $P_{5}^{'}$  of about $0.24$, which corresponds to approximately 34~\% of its Standard Model prediction. This shift is of the order of the deviation observed by the LHCb experiment.
    \item the solutions found by the RL Advocate and the $\chi^2$ minimisation provide very similar results. By definition, the RL solution cannot achieve values of $\chi^2$ lower than 0.1, since episodes are considered to be successfully terminated when the $\chi^2$ reaches the 0.1 threshold. The threshold of 0.1 is chosen as it is significantly smaller than one standard deviation and indeed decreasing this threshold to smaller values has a neglgible effect on the results.
    In general, it is sufficient to have a $\chi^2/\mathrm{d.o.f.}$ of the order of 1 to claim a good statistical description of a given series of measurements,
    however, due to the extremely simple example considered here, with only one input measurement, a threshold of 0.1 is chosen in order to get a solution that is as close as possible to the true minimum.
    \item the solutions obtained for case-I (trained only with the signal) would result in a shift on the  $P_{5}^{'}$ measurement for the control mode that is almost as large as the one on the signal, up to $\Delta {P_{5}^{'}}^{\jpsi} \simeq 0.15$, which is more than one order of magnitude larger than the uncertainty on the control channel measurement.
    \item when explicitly including the constraints from the control channel (case-II) we find that it is impossible for a mismodelling of the efficiency to cause a shift in the value of $P_{5}^{'}$ for $B^0 \to K^{*0} \mumu$ decays while keeping unchanged the measurement in the control channel, with the largest allowed variation of the order of 0.01.
\end{itemize}

\begin{table}[h]
\begin{center}
\caption{Solutions obtained by running the RL Advocate on the $P_{5}^{'}$ example.
The first two columns refer to the case where only the $B^0 \to K^{*0} \mumu$ measurement in $ q^2 \in [4.0, \, 6.0]$~GeV$^2$ is considered, with the first column (SciPy minimisation) showing the values obtained by running an analytic $\chi^2$ minimisation with the BFGS method from the SciPy python package~\cite{2020SciPy-NMeth}, while the second column (RL solution I) report the solution found by the RL Advocate algorithm.
The third column (RL solution II) shows the best solution obtained in presence of the constraint imposed by the control channel measurement.}
\label{tab:result_RL}
\begin{tabular}{@{}%
l%
S[table-format=3.3]%
S[table-format=3.3]%
S[table-format=3.3]%
@{}%
}
\toprule
{}
& {SciPy minimisation}
& {RL solution I}
& {RL solution II}
\\
\midrule
$k^+$             & 0.14  & 0.04      &   2.38   \\
$k^-$             & -1.58 & -1.09    &  0.38  \\
\midrule
$\Delta {P_{5}^{'}}^{sig}$      & 0.28 &  0.24   &   0.01    \\
$\Delta {P_{5}^{'}}^{\jpsi}$      &  0.14  &  0.15  &   0.0002  \\
\midrule
$\chi^2 / N_{\mathrm{meas}}$    & 0.0 &  0.1  &  5.3     \\
\bottomrule
\end{tabular}
\end{center}
\end{table}

\section{Discussion}
\label{sec:Discussion}

The two methods presented above demonstrate the ability to account for potentially hidden systematic uncertainties in measurements that may show a discrepancy between the values observed in the experiments and those predicted by the theory. 
While in this paper we employ simplified assumptions, which are necessary due to the impossibility to access realistic detector simulation and all the crosschecks performed during physics analyses that are not made public by the experiments, we set the ground for a novel methodology to access potentially hidden systematic uncertainties in high-energy physics measurements.
A final and more accurate implementation of the proposed method is therefore left to the individual experiments.

The first method relies on a combination of gradient descent and optimisation techniques and its application is illustrated with the example of a branching fraction measurement. This method can be extended to incorporate additional assumptions on the mismodelling of the efficiency which can be derived from physics considerations. 
As an example, we demonstrated how to impose a smoothening of the mismodelling function by adding an appropriate penalty term to the loss function.
These coarse assumptions can be avoided if low-level detector information is used rather than the high-level features used for this paper.

In principle, the first method can be extended to account for latent theoretical or detector parameters. 
Those parameters can become available within the simulation with the help of automatic differentiation~\cite{baydin2018automatic}. 
For example, one can explore cases where the measurement uncertainty depends on the detector's alignment or some coupling coefficient. However, most simulation packages are written in a way that doesn't support such techniques. This is precisely the motivation for studying RL-based approach, which can deal with non-differentiable function optimisations. The downside of this approach is that it is computationally more expensive compared to a classifier-based method. However, the flexibility afforded might be crucial in certain cases, particularly when dealing with effects difficult to parametrise. 
We illustrated the use of Reinforcement Learning with a simple example applied to the measurement of the  angular observable $P_{5}^{'}$ in $B^0 \to K^{*0} \mumu$ decays.
Despite the fact that the limited complexity of the considered example makes the examined case solvable by more standard approaches, we find that formulating the problem in terms of Reinforcement Learning can open a new avenue of Machine Learning applications in experimental particle physics.

In addition, throughout this paper we have only considered systematic effects from mismodelling of the efficiency as this is the simplest case to incorporate into the DL Advocate methodology. However, this can be extended to include several other types of systematic effects, such as signal resolution and backgrounds mismodellings or even theoretical uncertainties. Low-level information would again be very useful in order to generalise this approach to these issues.

Finally, with the growth of sensitivity of measuring devices and an inherent increase in the sensitivity to measured events, the degree of the uncertainty introduced during the measurement process becomes more and more eminent. Thus techniques helping estimating the boundaries of the observables that would confidently distinguish alternative physics hypotheses become an imminent research tool. Another reason pushing towards the development of such methods is that it becomes innately expensive to support a twin experiment to confirm expected discoveries, like CMS and ATLAS did for the discovery of the Higgs boson. The possible development of the Future Circular Collider~\cite{Abada_2019} would be rather difficult to match by any other research facility. Nevertheless, all the results and measurements should be scrutinised from the perspective of possibly unknown systematic uncertainties, which the devil's advocate approach is capable of. An interesting venue for exploring and extending this method would be the measurement of W mass, which has recently been measured by the CDF collaboration~\cite{doi:10.1126/science.abk1781} to be significantly different from previous measurements~\cite{CDF:2013dpa,ATLAS:2017rzl,LHCb:2021bjt,ATLAS-CONF-2023-004} and the Standard Model prediction~\cite{deBlas:2021wap}. This would require a large amount of numerical detail to be publicly released by experiments to have meaningful results such as auxillary inputs assumed and the exact variations that are applied in the measurements.

\section{Summary}

In summary, we have introduced a method to place quantitative bounds for hidden systematic effects using machine learning. The philosophy behind the approach is to play the devil's advocate by reversing the measurement process and assuming the Standard Model hypothesis, such that systematic nuisance parameters are determined by the measurements themselves. 

To demonstrate the method, we have applied the proposed approach to a hypothetical branching fraction measurement to see the maximum allowed bias due to possible mismodelling of the detector efficiency.
We find that the bias depends significantly on the kinematic overlap  between the signal and the control channels that are used as crosschecks in the analysis.

In addition to the nominal approach, a reinforcement learning technique is also employed.
We explore the potentiality of this alternative route by applying it to the measurement of the $P_{5}^{'}$ angular observable in $B^0 \to K^{*0} \mumu$ decays.
We find that in the tested minimal scenario, when including all the crosschecks of the analysis there is no room for hidden systematic uncertainties associated to the modelling of the detector efficiency to explain the deviation observed in data. 
This result is based on the use of fast simulation, a robust and final statement can only be achieved by considering all possible detector effects which can only be accessed with a full simulation of the detector.
The development of a reinforcement learning solution is an essential complement to the results of the paper, but such an approach can have the highest potential when generalised to low-level quantities.
The extension of the presented method to all possible aspects directly related to the detector, in fact, will be fundamental to demonstrate the robustness of any future discovery claim.


\addcontentsline{toc}{section}{References}
\bibliographystyle{JHEP}
\bibliography{references}

\providecommand{\href}[2]{#2}\begingroup\raggedright\begin{thebibliography}{10}

\bibitem{Aad_2012}
G.~Aad et~al., \emph{Observation of a new particle in the search for the
  standard model higgs boson with the {ATLAS} detector at the {LHC}},
  \href{https://doi.org/10.1016/j.physletb.2012.08.020}{\emph{Physics Letters
  B} {\bfseries 716} (2012) 1}
  [\href{https://arxiv.org/abs/1207.7214}{{\ttfamily 1207.7214}}].

\bibitem{Chatrchyan_2012}
S.~Chatrchyan et~al., \emph{Observation of a new boson at a mass of 125 {GeV}
  with the {CMS} experiment at the {LHC}},
  \href{https://doi.org/10.1016/j.physletb.2012.08.021}{\emph{Physics Letters
  B} {\bfseries 716} (2012) 30}
  [\href{https://arxiv.org/abs/1207.7235}{{\ttfamily 1207.7235}}].

\bibitem{Wunsch:2020iuh}
S.~Wunsch, S.~J\"orger, R.~Wolf and G.~Quast, \emph{{Optimal Statistical
  Inference in the Presence of Systematic Uncertainties Using Neural Network
  Optimization Based on Binned Poisson Likelihoods with Nuisance Parameters}},
  \href{https://doi.org/10.1007/s41781-020-00049-5}{\emph{Comput. Softw. Big
  Sci.} {\bfseries 5} (2021) 4}
  [\href{https://arxiv.org/abs/2003.07186}{{\ttfamily 2003.07186}}].

\bibitem{DeCastro:2018psv}
P.~De~Castro and T.~Dorigo, \emph{{INFERNO: Inference-Aware Neural
  Optimisation}},
  \href{https://doi.org/10.1016/j.cpc.2019.06.007}{\emph{Comput. Phys. Commun.}
  {\bfseries 244} (2019) 170}
  [\href{https://arxiv.org/abs/1806.04743}{{\ttfamily 1806.04743}}].

\bibitem{Ghosh:2021roe}
A.~Ghosh, B.~Nachman and D.~Whiteson, \emph{{Uncertainty-aware machine learning
  for high energy physics}},
  \href{https://doi.org/10.1103/PhysRevD.104.056026}{\emph{Phys. Rev. D}
  {\bfseries 104} (2021) 056026}
  [\href{https://arxiv.org/abs/2105.08742}{{\ttfamily 2105.08742}}].

\bibitem{Viren:2022qon}
B.~Viren, J.~Huang, Y.~Huang, M.~Lin, Y.~Ren, K.~Terao et~al., \emph{{Solving
  Simulation Systematics in and with AI/ML}},  in \emph{2022 Snowmass Summer
  Study}, 3, 2022,
  \href{https://arxiv.org/abs/2203.06112}{https://arxiv.org/abs/2203.06112}
  [\href{https://arxiv.org/abs/2203.06112}{{\ttfamily 2203.06112}}].

\bibitem{Englert:2018cfo}
C.~Englert, P.~Galler, P.~Harris and M.~Spannowsky, \emph{{Machine Learning
  Uncertainties with Adversarial Neural Networks}},
  \href{https://doi.org/10.1140/epjc/s10052-018-6511-8}{\emph{Eur. Phys. J. C}
  {\bfseries 79} (2019) 4} [\href{https://arxiv.org/abs/1807.08763}{{\ttfamily
  1807.08763}}].

\bibitem{Louppe:2016ylz}
G.~Louppe, M.~Kagan and K.~Cranmer, \emph{{Learning to Pivot with Adversarial
  Networks}},  \href{https://arxiv.org/abs/1611.01046}{{\ttfamily 1611.01046}}.

\bibitem{DescotesGenon:2012zf}
S.~Descotes-Genon, J.~Matias, M.~Ramon and J.~Virto, \emph{{Implications from
  clean observables for the binned analysis of $B \to K^*\mu^+\mu^-$ at large
  recoil}}, \href{https://doi.org/10.1007/JHEP01(2013)048}{\emph{JHEP}
  {\bfseries 01} (2013) 048} [\href{https://arxiv.org/abs/1207.2753}{{\ttfamily
  1207.2753}}].

\bibitem{Matias:2012xw}
J.~Matias, F.~Mescia, M.~Ramon and J.~Virto, \emph{{Complete Anatomy of
  $\bar{B}_d \to \bar{K}^{* 0} (\to K \pi)l^+l^-$ and its angular
  distribution}}, \href{https://doi.org/10.1007/JHEP04(2012)104}{\emph{JHEP}
  {\bfseries 04} (2012) 104} [\href{https://arxiv.org/abs/1202.4266}{{\ttfamily
  1202.4266}}].

\bibitem{doi:10.1126/science.abk1781}
T.~Aaltonen, S.~Amerio, D.~Amidei, A.~Anastassov, A.~Annovi, J.~Antos et~al.,
  \emph{{High-precision measurement of the $W$ boson mass with the CDF II
  detector}}, \href{https://doi.org/10.1126/science.abk1781}{\emph{Science}
  {\bfseries 376} (2022) 170}.

\bibitem{CDF:2013dpa}
{\scshape CDF, D0} collaboration, \emph{{Combination of CDF and D0 $W$-Boson
  Mass Measurements}},
  \href{https://doi.org/10.1103/PhysRevD.88.052018}{\emph{Phys. Rev. D}
  {\bfseries 88} (2013) 052018}
  [\href{https://arxiv.org/abs/1307.7627}{{\ttfamily 1307.7627}}].

\bibitem{ATLAS:2017rzl}
{\scshape ATLAS} collaboration, \emph{{Measurement of the $W$-boson mass in pp
  collisions at $\sqrt{s}=7$ TeV with the ATLAS detector}},
  \href{https://doi.org/10.1140/epjc/s10052-017-5475-4}{\emph{Eur. Phys. J. C}
  {\bfseries 78} (2018) 110}
  [\href{https://arxiv.org/abs/1701.07240}{{\ttfamily 1701.07240}}].

\bibitem{LHCb:2021bjt}
{\scshape LHCb} collaboration, \emph{{Measurement of the W boson mass}},
  \href{https://doi.org/10.1007/JHEP01(2022)036}{\emph{JHEP} {\bfseries 01}
  (2022) 036} [\href{https://arxiv.org/abs/2109.01113}{{\ttfamily
  2109.01113}}].

\bibitem{deBlas:2021wap}
J.~de~Blas, M.~Ciuchini, E.~Franco, A.~Goncalves, S.~Mishima, M.~Pierini
  et~al., \emph{Global analysis of electroweak data in the standard model},
  \href{https://doi.org/10.1103/PhysRevD.106.033003}{\emph{Phys. Rev. D}
  {\bfseries 106} (2022) 033003}
  [\href{https://arxiv.org/abs/2112.07274}{{\ttfamily 2112.07274}}].

\bibitem{2020SciPy-NMeth}
P.~Virtanen, R.~Gommers, T.E.~Oliphant, M.~Haberland, T.~Reddy, D.~Cournapeau
  et~al., \emph{{{SciPy} 1.0: Fundamental Algorithms for Scientific Computing
  in Python}}, \href{https://doi.org/10.1038/s41592-019-0686-2}{\emph{Nature
  Methods} {\bfseries 17} (2020) 261}.

\bibitem{matrixcookbook}
K.B.~Petersen and M.S.~Pedersen, \emph{The matrix cookbook},  11, 2012.

\bibitem{PDG:2022pth}
{\scshape Particle Data Group} collaboration, \emph{{Review of Particle
  Physics}}, \href{https://doi.org/10.1093/ptep/ptac097}{\emph{PTEP} {\bfseries
  2022} (2022) 083C01}.

\bibitem{LHCb:2016ehk}
{\scshape LHCb} collaboration, \emph{{Measurement of the ratio of branching
  fractions and difference in $CP$ asymmetries of the decays $B^+\to
  J/\psi\pi^+$ and $B^+\to J/\psi K^+$}},
  \href{https://doi.org/10.1007/JHEP03(2017)036}{\emph{JHEP} {\bfseries 03}
  (2017) 036} [\href{https://arxiv.org/abs/1612.06116}{{\ttfamily
  1612.06116}}].

\bibitem{LHCb:2008vvz}
{\scshape LHCb} collaboration, \emph{{The LHCb Detector at the LHC}},
  \href{https://doi.org/10.1088/1748-0221/3/08/S08005}{\emph{JINST} {\bfseries
  3} (2008) S08005}.

\bibitem{Cowan:2016tnm}
G.A.~Cowan, D.C.~Craik and M.D.~Needham, \emph{{RapidSim: an application for
  the fast simulation of heavy-quark hadron decays}},
  \href{https://doi.org/10.1016/j.cpc.2017.01.029}{\emph{Comput. Phys. Commun.}
  {\bfseries 214} (2017) 239}
  [\href{https://arxiv.org/abs/1612.07489}{{\ttfamily 1612.07489}}].

\bibitem{Davidson:2010ew}
N.~Davidson, T.~Przedzinski and Z.~Was, \emph{{PHOTOS interface in C++:
  Technical and Physics Documentation}},
  \href{https://doi.org/10.1016/j.cpc.2015.09.013}{\emph{Comput. Phys. Commun.}
  {\bfseries 199} (2016) 86} [\href{https://arxiv.org/abs/1011.0937}{{\ttfamily
  1011.0937}}].

\bibitem{LHCb:2014cxe}
{\scshape LHCb} collaboration, \emph{{Differential branching fractions and
  isospin asymmetries of $B \to K^{(*)} \mu^+ \mu^-$ decays}},
  \href{https://doi.org/10.1007/JHEP06(2014)133}{\emph{JHEP} {\bfseries 06}
  (2014) 133} [\href{https://arxiv.org/abs/1403.8044}{{\ttfamily 1403.8044}}].

\bibitem{sutton2018reinforcement}
R.S.~Sutton and A.G.~Barto, \emph{Reinforcement learning: An introduction}, MIT
  press, Cambridge, MA (2018).

\bibitem{li2018deep}
Y.~Li, \emph{Deep reinforcement learning},
  \href{https://arxiv.org/abs/1810.06339}{{\ttfamily 1810.06339}}.

\bibitem{kalashnikov2018qtopt}
D.~Kalashnikov, A.~Irpan, P.~Pastor, J.~Ibarz, A.~Herzog, E.~Jang et~al.,
  \emph{Qt-opt: Scalable deep reinforcement learning for vision-based robotic
  manipulation},  \href{https://arxiv.org/abs/1806.10293}{{\ttfamily
  1806.10293}}.

\bibitem{kiran2021deep}
B.R.~Kiran, I.~Sobh, V.~Talpaert, P.~Mannion, A.A.A.~Sallab, S.~Yogamani
  et~al., \emph{Deep reinforcement learning for autonomous driving: A survey},
  \href{https://arxiv.org/abs/2002.00444}{{\ttfamily 2002.00444}}.

\bibitem{silver2017mastering}
D.~Silver, J.~Schrittwieser, K.~Simonyan, I.~Antonoglou, A.~Huang, A.~Guez
  et~al., \emph{Mastering the game of go without human knowledge},
  \href{https://doi.org/10.1038/nature24270}{\emph{Nature} {\bfseries 550}
  (2017) 354}.

\bibitem{Carrazza:2019efs}
S.~Carrazza and F.A.~Dreyer, \emph{{Jet grooming through reinforcement
  learning}}, \href{https://doi.org/10.1103/PhysRevD.100.014014}{\emph{Phys.
  Rev. D} {\bfseries 100} (2019) 014014}
  [\href{https://arxiv.org/abs/1903.09644}{{\ttfamily 1903.09644}}].

\bibitem{Cranmer:2021gdt}
K.~Cranmer, M.~Drnevich, S.~Macaluso and D.~Pappadopulo, \emph{{Reframing Jet
  Physics with New Computational Methods}},
  \href{https://doi.org/10.1051/epjconf/202125103059}{\emph{EPJ Web Conf.}
  {\bfseries 251} (2021) 03059}
  [\href{https://arxiv.org/abs/2105.10512}{{\ttfamily 2105.10512}}].

\bibitem{StJohn:2020bpk}
J.~St.~John et~al., \emph{{Real-time artificial intelligence for accelerator
  control: A study at the Fermilab Booster}},
  \href{https://doi.org/10.1103/PhysRevAccelBeams.24.104601}{\emph{Phys. Rev.
  Accel. Beams} {\bfseries 24} (2021) 104601}
  [\href{https://arxiv.org/abs/2011.07371}{{\ttfamily 2011.07371}}].

\bibitem{liang2018rllib}
E.~Liang, R.~Liaw, P.~Moritz, R.~Nishihara, R.~Fox, K.~Goldberg et~al.,
  \emph{Rllib: Abstractions for distributed reinforcement learning},
  \href{https://arxiv.org/abs/1712.09381}{{\ttfamily 1712.09381}}.

\bibitem{DQN}
V.~Mnih, K.~Kavukcuoglu, D.~Silver, A.~Graves, I.~Antonoglou, D.~Wierstra
  et~al., \emph{Playing atari with deep reinforcement learning},
  \href{https://arxiv.org/abs/1312.5602}{{\ttfamily 1312.5602}}.

\bibitem{DoubleDQN}
H.~van Hasselt, A.~Guez and D.~Silver, \emph{Deep reinforcement learning with
  double q-learning},  \href{https://arxiv.org/abs/1509.06461}{{\ttfamily
  1509.06461}}.

\bibitem{DuelingDQN}
Z.~Wang, T.~Schaul, M.~Hessel, H.~van Hasselt, M.~Lanctot and N.~de~Freitas,
  \emph{Dueling network architectures for deep reinforcement learning},
  \href{https://arxiv.org/abs/1511.06581}{{\ttfamily 1511.06581}}.

\bibitem{Altmannshofer:2013foa}
W.~Altmannshofer and D.M.~Straub, \emph{{New Physics in $B \to K^*\mu\mu$?}},
  \href{https://doi.org/10.1140/epjc/s10052-013-2646-9}{\emph{Eur. Phys. J.}
  {\bfseries C73} (2013) 2646}
  [\href{https://arxiv.org/abs/1308.1501}{{\ttfamily 1308.1501}}].

\bibitem{Gauld:2013qja}
R.~Gauld, F.~Goertz and U.~Haisch, \emph{{An explicit Z'-boson explanation of
  the $B \to K^* \mu^+ \mu^-$ anomaly}},
  \href{https://doi.org/10.1007/JHEP01(2014)069}{\emph{JHEP} {\bfseries 01}
  (2014) 069} [\href{https://arxiv.org/abs/1310.1082}{{\ttfamily 1310.1082}}].

\bibitem{Buras:2013qja}
A.J.~Buras and J.~Girrbach, \emph{{Left-handed $Z'$ and $Z$ FCNC quark
  couplings facing new $b \to s \mu^+ \mu^-$ data}},
  \href{https://doi.org/10.1007/JHEP12(2013)009}{\emph{JHEP} {\bfseries 12}
  (2013) 009} [\href{https://arxiv.org/abs/1309.2466}{{\ttfamily 1309.2466}}].

\bibitem{Crivellin:2015era}
A.~Crivellin, L.~Hofer, J.~Matias, U.~Nierste, S.~Pokorski and J.~Rosiek,
  \emph{{Lepton-flavour violating $B$ decays in generic $Z'$ models}},
  \href{https://doi.org/10.1103/PhysRevD.92.054013}{\emph{Phys. Rev.}
  {\bfseries D92} (2015) 054013}
  [\href{https://arxiv.org/abs/1504.07928}{{\ttfamily 1504.07928}}].

\bibitem{LHCb:2020lmf}
{\scshape LHCb} collaboration, \emph{{Measurement of $CP$-Averaged Observables
  in the $B^{0}\rightarrow K^{*0}\mu^{+}\mu^{-}$ Decay}},
  \href{https://doi.org/10.1103/PhysRevLett.125.011802}{\emph{Phys. Rev. Lett.}
  {\bfseries 125} (2020) 011802}
  [\href{https://arxiv.org/abs/2003.04831}{{\ttfamily 2003.04831}}].

\bibitem{Wehle:2016yoi}
{\scshape Belle} collaboration, \emph{{Lepton-Flavor-Dependent Angular Analysis
  of $B\to K^\ast \ell^+\ell^-$}},
  \href{https://doi.org/10.1103/PhysRevLett.118.111801}{\emph{Phys. Rev. Lett.}
  {\bfseries 118} (2017) 111801}
  [\href{https://arxiv.org/abs/1612.05014}{{\ttfamily 1612.05014}}].

\bibitem{ATLAS:2018gqc}
{\scshape ATLAS} collaboration, \emph{{Angular analysis of $B^0_d \rightarrow
  K^{*}\mu^+\mu^-$ decays in $pp$ collisions at $\sqrt{s}= 8$ TeV with the
  ATLAS detector}}, \href{https://doi.org/10.1007/JHEP10(2018)047}{\emph{JHEP}
  {\bfseries 10} (2018) 047}
  [\href{https://arxiv.org/abs/1805.04000}{{\ttfamily 1805.04000}}].

\bibitem{CMS:2017rzx}
{\scshape CMS} collaboration, \emph{{Measurement of angular parameters from the
  decay $\mathrm{B}^0 \to \mathrm{K}^{*0} \mu^+ \mu^-$ in proton-proton
  collisions at $\sqrt{s} = $ 8 TeV}},
  \href{https://doi.org/10.1016/j.physletb.2018.04.030}{\emph{Phys. Lett. B}
  {\bfseries 781} (2018) 517}
  [\href{https://arxiv.org/abs/1710.02846}{{\ttfamily 1710.02846}}].

\bibitem{Bharucha:2015bzk}
A.~Bharucha, D.M.~Straub and R.~Zwicky, \emph{{$B\to V\ell^+\ell^-$ in the
  Standard Model from light-cone sum rules}},
  \href{https://doi.org/10.1007/JHEP08(2016)098}{\emph{JHEP} {\bfseries 08}
  (2016) 098} [\href{https://arxiv.org/abs/1503.05534}{{\ttfamily
  1503.05534}}].

\bibitem{Altmannshofer:2014rta}
W.~Altmannshofer and D.M.~Straub, \emph{{New physics in $b\rightarrow s$
  transitions after LHC run 1}},
  \href{https://doi.org/10.1140/epjc/s10052-015-3602-7}{\emph{Eur. Phys. J. C}
  {\bfseries 75} (2015) 382} [\href{https://arxiv.org/abs/1411.3161}{{\ttfamily
  1411.3161}}].

\bibitem{Khodjamirian:2010vf}
A.~Khodjamirian, T.~Mannel, A.A.~Pivovarov and Y.M.~Wang, \emph{{Charm-loop
  effect in $B \to K^{(*)} \ell^{+} \ell^{-}$ and $B\to K^*\gamma$}},
  \href{https://doi.org/10.1007/JHEP09(2010)089}{\emph{JHEP} {\bfseries 09}
  (2010) 089} [\href{https://arxiv.org/abs/1006.4945}{{\ttfamily 1006.4945}}].

\bibitem{Descotes-Genon:2014uoa}
S.~Descotes-Genon, L.~Hofer, J.~Matias and J.~Virto, \emph{{On the impact of
  power corrections in the prediction of $B \to K^*\mu^+\mu^-$ observables}},
  \href{https://doi.org/10.1007/JHEP12(2014)125}{\emph{JHEP} {\bfseries 12}
  (2014) 125} [\href{https://arxiv.org/abs/1407.8526}{{\ttfamily 1407.8526}}].

\bibitem{LHCb:2015svh}
{\scshape LHCb} collaboration, \emph{{Angular analysis of the $B^{0} \to K^{*0}
  \mu^{+} \mu^{-}$ decay using 3 fb$^{-1}$ of integrated luminosity}},
  \href{https://doi.org/10.1007/JHEP02(2016)104}{\emph{JHEP} {\bfseries 02}
  (2016) 104} [\href{https://arxiv.org/abs/1512.04442}{{\ttfamily
  1512.04442}}].

\bibitem{Altmannshofer:2008dz}
W.~Altmannshofer, P.~Ball, A.~Bharucha, A.J.~Buras, D.M.~Straub and M.~Wick,
  \emph{{Symmetries and Asymmetries of $B \to K^{*} \mu^{+} \mu^{-}$ Decays in
  the Standard Model and Beyond}},
  \href{https://doi.org/10.1088/1126-6708/2009/01/019}{\emph{JHEP} {\bfseries
  01} (2009) 019} [\href{https://arxiv.org/abs/0811.1214}{{\ttfamily
  0811.1214}}].

\bibitem{LHCb:2014xzf}
{\scshape LHCb} collaboration, \emph{{Measurement of polarization amplitudes
  and CP asymmetries in $B^0 \to \phi K^*(892)^0$}},
  \href{https://doi.org/10.1007/JHEP05(2014)069}{\emph{JHEP} {\bfseries 05}
  (2014) 069} [\href{https://arxiv.org/abs/1403.2888}{{\ttfamily 1403.2888}}].

\bibitem{Belle:2014nuw}
{\scshape Belle} collaboration, \emph{{Observation of a new charged
  charmoniumlike state in $\bar{B}^0 \to J\!/\!\psi K^-\pi^+$ decays}},
  \href{https://doi.org/10.1103/PhysRevD.90.112009}{\emph{Phys. Rev. D}
  {\bfseries 90} (2014) 112009}
  [\href{https://arxiv.org/abs/1408.6457}{{\ttfamily 1408.6457}}].

\bibitem{baydin2018automatic}
A.G.~Baydin, B.A.~Pearlmutter, A.A.~Radul and J.M.~Siskind, \emph{Automatic
  differentiation in machine learning: a survey}, {\emph{Journal of Marchine
  Learning Research} {\bfseries 18} (2018) 5595–}
  [\href{https://arxiv.org/abs/1502.05767}{{\ttfamily 1502.05767}}].

\bibitem{Abada_2019}
A.~Abada et~al., \emph{{FCC} physics opportunities},
  \href{https://doi.org/10.1140/epjc/s10052-019-6904-3}{\emph{Eur. Phys. J. C}
  {\bfseries 79} (2019) }.

\bibitem{ATLAS-CONF-2023-004}
{\scshape ATLAS} collaboration, \emph{{Improved W boson Mass Measurement using
  7 TeV Proton-Proton Collisions with the ATLAS Detector}}, .

\bibitem{pytorch}
A.~Paszke, S.~Gross, F.~Massa, A.~Lerer, J.~Bradbury, G.~Chanan et~al.,
  \emph{Pytorch: An imperative style, high-performance deep learning library},
  2019.

\end{thebibliography}\endgroup


\appendix

\numberwithin{table}{section}
\numberwithin{figure}{section}
\numberwithin{algorithm}{section}

\clearpage
\section{DL Advocate}\label{app:algorithm}

The formal definition of the optimization algorithm is presented in Algorithm~\ref{alg:training}.

\begin{algorithm}
\caption{DL Advocate: linear programming optimisation.}\label{alg:training}
\begin{algorithmic}[1]
\Require{dataset $D=\{(x_i,y_i)\}_{i=1}^N$, efficiency bounds $B=\{(e_j^l, e_j^u)\}_{j=1}^k$, NN $f(x;\theta)$}
\Require{Learning rate $\eta$, gradient penalty $p$, numeric differentiation step $\xi$}
\State{$h(x;\theta) \equiv \operatorname{softmax}(f(x;\theta))$}
\State{$\theta\gets\arg\min_\theta\sum_{(x,y)\in D}\operatorname{crossentropy loss}(y,h(x;\theta))$}
\While{not converged}
    \State{$\partial\ell \gets 0$}
    \ForAll{$i, y \in 1 \dots k$}
    \State{$H_{i,y} \gets \frac{1}{N_y}\sum_{j: y_j=y}h_i(x_j;\theta)$}
    \EndFor
    \State{$H^+ \gets \operatorname{inverse}(H)$}
    \State{$\alpha \gets \operatorname{solve LP}(H, B)$}
    \State{$E \gets \operatorname{estimator}(H, B, \alpha)$}
    \For{$(x,y) \in D$}
    \State{$\partial\ell_{sd} \gets \frac{1}{N_y}\sum_{i=1}^k(E_{y,i}-H^+_{y,i}) \frac{\partial h_i(x;\theta)}{\partial\theta}$}
    \State{$\partial\ell \gets \partial\ell + \partial\ell_{sd}$}
    \If{GP is enabled}
        \ForAll{$i \in 1 \dots k$}
        \State{$\partial x_i \gets \frac\partial{\partial x}h_i(x;\theta)$}
        \State{$L_i \gets \frac{h_i(x+\xi\partial x_i;\theta)-h_i(x-\xi\partial x_i;\theta))}{2\xi\left\Vert \partial x_i\right\Vert}$}
        \EndFor
        \State{$\partial\ell_g \gets \frac1N \frac\partial{\partial\theta}\left(\frac1k\sum_{i=1}^k\left(\frac{L_i}{p}-1\right)^2\right)$}
        \State{$\partial\ell \gets \partial\ell + \partial\ell_g$}
    \EndIf
    \EndFor
    \State{$\theta\gets\theta - \eta \partial\ell$}
\EndWhile
\ForAll{$i, y \in 1 \dots k$}
\State{$H_{i,y} \gets \frac{1}{N_y}\sum_{j: y_j=y}h_i(x_j;\theta)$}
\EndFor
\State{$\alpha \gets \operatorname{solve LP}(H, B)$}
\State{\Return{$\alpha, \theta$}}
\end{algorithmic}
\end{algorithm}

Training procedure is implemented using PyTorch framework \cite{pytorch}.
The neural network $f(x;\theta)$ is composed of three internal linear layers
with 20 outputs each and $\operatorname{softplus}$ activations.
The final linear layer has three output nodes, which correspond to the number of considered channels.
This network is rather simple but it is sufficient to represent
non-trivial interpretable weightings. Softplus activations are used
to keep the resulting function smooth in the domain.
Input data is normalized before training to have same scale between features, while the gradient penalty term is computed after feature normalization.

\end{document}